\documentclass[12pt]{iopart}
\usepackage{iopams}
\usepackage{graphicx}

\begin{document}
 
\title[Cosmic Topology of Prism Double-Action Manifolds]
{Cosmic Topology of Prism Double-Action Manifolds}
\author{R.\ Aurich and S.\ Lustig}
 
\address{Institut f\"ur Theoretische Physik, Universit\"at Ulm,\\
Albert-Einstein-Allee 11, D-89069 Ulm, Germany}

\begin{abstract}
The cosmic microwave background (CMB) anisotropies in spherical 3-spaces
with a non-trivial topology are studied.
This paper discusses the special class of the so-called double-action
manifolds, which are for the first time analysed with respect
to their CMB anisotropies.
The CMB anisotropies are computed for all prism double-action manifolds
generated by a binary dihedral and a cyclic group
with a group order of up to 180 leading to 33 different topologies.
Several spaces are found which show a suppression of the CMB anisotropies
on large angular distances as it is found on the real CMB sky.
It turns out that two of these spaces possess Dirichlet domains
which are not very far from highly symmetric
polyhedra like Platonic or Archimedean ones.
\end{abstract}


\submitto{\CQG}

\section{Introduction.}
\label{sec:intro}

The NASA satellite COBE was not only the first mission which discovers
fluctuations in the cosmic microwave background (CMB) radiation,
but it also revealed that these fluctuations
are almost uncorrelated at large angular scales
\cite{Hinshaw_et_al_1996}.
This important observation was later substantiated by the
WMAP mission \cite{Spergel_et_al_2003},
and it is described by the temperature 2-point correlation function
\begin{equation}
\label{Eq:C_theta}
C(\vartheta) \; := \; \left< \delta T(\hat n) \delta T(\hat n')\right>
\hspace{10pt} \hbox{with} \hspace{10pt}
\hat n \cdot \hat n' = \cos\vartheta
\hspace{10pt} ,
\end{equation}
where $\delta T(\hat n)$ is the temperature fluctuation in
the direction of the unit vector $\hat n$.
Since the correlations are most strongly suppressed at
angles $\vartheta\gtrsim 60^\circ$,
the scalar measure
\begin{equation}
\label{Eq:S_statistic_60}
S \; := \; \int^{\cos(60^\circ)}_{\cos(180^\circ)}
\hbox{d}\cos\vartheta \; |C(\vartheta)|^2
\hspace{10pt}
\end{equation}
is introduced in \cite{Spergel_et_al_2003}.
Small values of the $S$ statistics signify a low correlation at large angles.

The observed low values of the $S$ statistics cannot be easily
reconciled with the cosmological $\Lambda$CDM concordance model.
Among the suggested possibilities to explain this behaviour is
that the spatial space might not be infinite as assumed
by the concordance model.
The space could possess a non-trivial topology
which can lead to multiconnected spaces with a finite volume.
Due to the lower cutoff in their wavenumber spectrum $\{k\}$,
these spaces can naturally explain the low correlations in the CMB sky.
More details on the cosmic topology can be found in
\cite{Lachieze-Rey_Luminet_1995,Luminet_Roukema_1999,Levin_2002,%
Reboucas_Gomero_2004,Luminet_2008}.

In this paper, it is assumed that the spatial space has a slight
positive curvature, so that the simply connected space is the
spherical 3-space ${\cal S}^3$ which can be embedded in
the four-dimensional Euclidean space as a 3-sphere
$$
\vec{x} \; = \; (x_0,x_1,x_2,x_3)^T\in {\cal S}^3
$$
together with the constraint $|\vec x\,|^2 = x_0^2+x_1^2+x_2^2+x_3^2 = 1$.
Using complex coordinates $z_1 := x_0 + \hbox{i} x_3$ and
$z_2 := x_1 - \hbox{i} x_2$,
the coordinate matrix $u$ can be defined 
\begin{equation}
\label{Eq:coordinates_u}
u \; := \;
\left(\begin{array}{cc}
z_1 & \hbox{i} z_2 \\ \hbox{i} \overline{z}_2 & \overline{z}_1
\end{array}\right)
\in \hbox{SU}(2,\mathbb{C}) \equiv {\cal S}^3
\hspace{10pt} .
\end{equation}
The advantage of the complex representation is that the transformations
on ${\cal S}^3$ are determined by two $\hbox{SU}(2,\mathbb {C})$ matrices
denoted as the pair $(g_a, g_b)$
that acts on the points $u \in \hbox{SU}(2,\mathbb{C})$ of the 3-sphere
${\cal S}^3 \equiv \hbox{SU}(2,\mathbb{C})$ by left and right multiplication
\begin{equation}
\label{Ref:act}
g \, := \, (g_a, g_b): \, u \rightarrow g_a^{-1}\,u\, g_b
\hspace{10pt} .
\end{equation}
The points $u$ and $g(u)$ are identified
if $g$ belongs to a deck group $\Gamma$.
The 3-sphere ${\cal S}^3$ is tessellated in this way by a deck group $\Gamma$
into as many domains as the deck group has elements
that is the order $|\Gamma|$ of the deck group.
The deck groups that lead to spherical multiconnected manifolds
are discussed in \cite{Gausmann_Lehoucq_Luminet_Uzan_Weeks_2001}.

In the following the focus is put on the double-action manifolds
that are generated by two finite subgroups $R$ and $L$ of Clifford translations.
The defining property of Clifford translations is that all points $u$
are translated by the same spherical distance.
Furthermore, they can be divided into left- and right-handed
Clifford translations depending on whether the flow lines
spiral clockwise and anticlockwise around each other, respectively.
In order to obtain a fix-point free group $\Gamma$,
the two subgroups $R$ and $L$ have to fulfil some conditions
\cite{Gausmann_Lehoucq_Luminet_Uzan_Weeks_2001}.
It turns out that either $R$ or $L$ must be cyclic, and
we take $L=Z_n$ as the cyclic subgroup of Clifford translations
without loss of generality.
The subgroup $R$ is chosen as the binary dihedral group $D^\star_p$
where the group orders $n$ and $p$ must not have a common
divisor greater than one. 

The cyclic subgroup $L=Z_n$ of Clifford translations
is generated by
\begin{equation}
\label{Def:L_p_q}
g_{l1} \; = \; ({\bf 1},g_b)
\hspace{10pt} \hbox{ with }\hspace{10pt}
g_b \; = \;
\hbox{diag}(e^{+2\pi\hbox{\scriptsize i}/n},e^{-2\pi\hbox{\scriptsize i}/n})
\hspace{10pt} ,
\end{equation}
since left-handed Clifford translations are realised by right multiplication.
The other elements of $L$ are obtained from (\ref{Def:L_p_q}) by
\begin{equation}
\label{Def:L_group}
g_{lk} \; = \; \Big( {\bf 1}, (g_b)^k \Big)
\hspace{10pt} \hbox{ for } \hspace{10pt}
k = 1, \dots, n
\hspace{10pt} .
\end{equation}
The binary dihedral group $R = D^\star_p$ has the two generators
$g_{r1}=(g_{a1 },{\bf 1})$ and $g_{r2}=(g_{a2},{\bf 1})$ with
\begin{equation}
\label{Def:D_p}
g_{a1} \; = \;
\hbox{diag}(e^{-\hbox{\scriptsize i}\Psi_{az}},e^{\hbox{\scriptsize i}\Psi_{az}})
\hspace{10pt} \hbox{ and } \hspace{10pt}
g_{a2} \; = \;
\left(\begin{array}{cc}
\cos(\Psi_{ay}) & -\sin(\Psi_{ay})\\ \sin(\Psi_{ay})& \cos(\Psi_{ay}) 
\end{array}\right)
\hspace{6pt} ,
\end{equation}
where $\Psi_{az}=2\pi\left(\frac{2}p\right)$ and
$\Psi_{ay}=2\pi\left(\frac{1}4\right)$.
The deck group $\Gamma$ consists of all combinations of
the elements of the subgroups $R$ and $L$.
The group order $|\Gamma|$ of the prism double-action deck group is
thus $|\Gamma|=n\,p$.
In the following, the manifold generated by $\Gamma$ is denoted as
$DZ(p,n) := {\cal S}^3/\left(D^\star_p \times Z_n\right)$
where the letters $D$ and $Z$ in $DZ(p,n)$ indicate the type of the group
that is the binary dihedral and the cyclic group
with group orders $p$ and $n$, respectively.

\section{Transforming the CMB Observer in Double-Action Manifolds}
\label{sec:obsv_trafo}

Before we can proceed to the analysis of the CMB anisotropies
of double-action manifolds,
we have to consider the transformation of the position
$u \in \hbox{SU}(2,\mathbb{C}) \equiv {\cal S}^3$ of the observer
for which the CMB anisotropy is to be computed.
Interestingly, it turns out that the statistical CMB properties depend
on the observer position within the double-action manifold.

The transformation is realised by applying an arbitrary isometry $t$
to the coordinates
\begin{equation} 
\label{Eq:trafo_coordinate}
u \rightarrow u'= u\, t \; \; , \; \; t \in \hbox{SU}(2,\mathbb{C})
\hspace{10pt} .
\end{equation}
Thus, the transformation is defined as right multiplication.
Interpreting such a transformation as a shift of the origin of
the coordinate system leads to a new representation of
the group elements $g_k = (g_{ak},g_{bk})$ of the deck group as
\begin{equation}
\label{Eq:trafo_group}
g'_k \; = \; (g'_{ak},g'_{bk}) \; = \; (g_{ak},t^{-1}\,g_{bk}\,t)
\hspace{10pt} , \hspace{10pt}
k=1, 2,\dots,|\Gamma|
\hspace{10pt} ,
\end{equation}
as shown in \cite{Aurich_Kramer_Lustig_2011,Aurich_Lustig_2012a}.
Thus, only the elements of the subgroup $L$ are altered.
The position dependence can conveniently be described by using
the parameterisation
\begin{equation}
\label{Eq:coordinate_t_rho_alpha_epsilon}
t(\rho,\alpha,\epsilon) \; = \;
\left( \begin{array}{cc}
\cos(\rho)\, e^{+\hbox{\scriptsize i}\alpha} &
\sin(\rho)\, e^{+\hbox{\scriptsize i}\epsilon} \\
-\sin(\rho)\, e^{-\hbox{\scriptsize i}\epsilon} &
\cos(\rho)\, e^{-\hbox{\scriptsize i}\alpha}
\end{array} \right)
\end{equation} 
for the transformation matrix $t$ with
$\rho \in [0,\frac{\pi}2]$, $\alpha, \epsilon \in [0,2\pi]$.
In this way the group elements $g_k$ of $\Gamma$ are functions of the
parameters $\rho$, $\alpha$, and $\epsilon$.

\section{Eigenmodes in Double-Action Manifolds}
\label{sec:eigenmode}

Whether the dependence of the group elements $g\in\Gamma$ on the
three parameters $\rho$, $\alpha$, and $\epsilon$ carries over to
the correlations in the CMB anisotropy
or only a subset of the parameters, is an interesting property of the
deck group $\Gamma$ and of the multiconnected manifold.
The statistical measures of the CMB are constructed rotationally invariant
in order to obtain a measure
that does not depend on the orientation of the coordinate system.
Thus, if the transformation (\ref{Eq:trafo_group}) leads to
new group elements that can be considered as a pure rotation of
the old ones, the statistical properties of the CMB do not change.

For any two points $P$ and $Q$ in a homogeneous manifold,
there exists a global isometry of the manifold taking $P$ to $Q$.
This implies that the CMB statistics do not depend on the observer position
parameterised by $\rho$, $\alpha$, and $\epsilon$ in this case.
For inhomogeneous manifolds there is no such global isometry and
the statistical properties of the CMB depend on the observer position,
in general.
Therefore, the CMB analysis requires a much more detailed investigation.
In the case of the inhomogeneous lens spaces $L(p,q)$
which are studied in \cite{Aurich_Lustig_2012b},
there is only a $\rho$ dependence, and the position dependent
CMB analysis reduces to a one dimensional scan.
This applies also to those inhomogeneous lens spaces
which are coincidentally double-action manifolds.
It turns out, however, that the ensemble averages of the CMB properties
of the double-action manifolds $DZ(p,n)$
depend on two parameters, for which $\rho$ and $\alpha$ are chosen.
To show this, we need to discuss the Laplace-Beltrami operator $\Delta$.

The eigenmodes of the Laplace-Beltrami operator $\Delta$ on the 
simply connected spherical manifold ${\cal S}^3$ can be given as a product 
of the eigenmodes $|j_a,m_a\rangle$ and $|j_b,m_b\rangle$
of the abstract generators of the Lie algebra
$\vec J_a = (J_{ax},J_{ay},J_{az}) \in \hbox{SU}_a(2,\mathbb{C})$
and
$\vec J_b = (J_{bx},J_{by},J_{bz}) \in \hbox{SU}_b(2,\mathbb{C})$,
respectively, for more details see \cite{Aurich_Lustig_2012a}.
Then, the complete set of eigenmodes for the eigenvalue 
$E_j := 4j(j+1) = (\beta^2-1)$  of the operator $-\Delta$ is obtained by
\begin{equation}
\label{Eq:SO4_Basis}
| j; m_a, m_b \rangle \; := \; |j,m_a\rangle \, |j,m_b\rangle \in \hbox{SO}(4,\mathbb{R})
\hspace{10pt} .
\end{equation}
In this notation the action of the generator (\ref{Def:L_p_q}) 
of the cyclic group $Z_n$ is described by 
$U_{g_{l1}}=e^{\hbox{\scriptsize i} (4\pi/n) J_{bz}}$.
Analogously the action of the two generators (\ref{Def:D_p}) 
of the binary dihedral group $D^\star_p$ are given by 
$U_{g_{r1}}=e^{\hbox{\scriptsize i} 2\psi_{az}J_{az}}$ and 
$U_{g_{r2}}=e^{\hbox{\scriptsize i} 2\psi_{ay}J_{ay}}$ 
with $\psi_{az}$ and $\psi_{ay}$ defined below eq.\,(\ref{Def:D_p}).
The eigenmodes on the manifold $DZ(p,n)$
have to be invariant under the action of $U_{g_{l1}}$, $U_{g_{r1}}$, and $U_{g_{r2}}$
\cite{Lustig_2007}, which is satisfied by
\begin{eqnarray}
\label{Eq:eigenmodes_DZ}
|j,i \rangle=\left\{
\begin{array}{ll}
|j; m_a, m_b \rangle &
: \;  m_a=0 \hbox{ for } j \hbox{ even} \\
\frac{1}{\sqrt{2}}\left(|j; m_a, m_b \rangle+(-1)^{j+m_a}\,|j; -m_a, m_b \rangle\right) &
: m_a>0
\end{array}\right.
\hspace{10pt}
\end{eqnarray}
where $j\in  \mathbb{N}_0 \setminus \{1,3,...,2 [\frac{p}{8}]-1\}$, $m_b \in \mathbb{Z}$,
$m_a \in \mathbb{N}_0$, $m_a\equiv 0\;\hbox{mod}\;p/4$, $2\,m_b\equiv 0\;\hbox{mod}\;n$
and $m_a,\left|m_b\right|\le j$.
The index $i$ with $1 \le i \le r^{DZ(p,n)}(\beta)$
counts the degenerated states which belong to the same eigenvalue $E_j$.
The indices $m_a$ and $m_b$ can be considered as functions of the
degeneracy index $i$, i.\,e.\ $m_a=m_a(i)$ and $m_b=m_b(i)$.
An analytic expression for the multiplicity $r^{DZ(p,n)}(\beta)$ is stated
in  table \ref{Tab:Spectrum}.
This table also gives the multiplicity of the eigenvalue $E_j$
for the double-action manifolds 
$TZ(24,n)={\cal S}^3/\left(T^\star \times Z_n\right)$,
$OZ(48,n)={\cal S}^3/\left(O^\star \times Z_n\right)$, and
$IZ(120,n)={\cal S}^3/\left(I^\star \times Z_n\right)$, 
where $T^\star$, $O^\star$, and $I^\star$ are the binary tetrahedral,
the binary octahedral, and the binary icosahedral group.
The difference between these formulae of the double-action manifolds and
the expressions for the multiplicity of the corresponding
homogeneous manifolds, 
see e.\,g.\ table 1 in \cite{Aurich_Lustig_2012a}, results from
the additional constraint $2\,m_b\equiv 0\;\hbox{mod}\;n$ due to
the cyclic group $Z_n$.
Therefore, the formulae for the multiplicity of the homogeneous manifolds 
are reproduced for $n=1$.  
 
\begin{table}[!htbp]
\centering
 \hspace*{-0.5cm}\begin{tabular}{|c|c|c|}
 \hline
 manifold ${\cal M}$ & wave number spectrum $\{\beta\}$  &
 multiplicity $r^{\cal M}(\beta)$ \\
 \hline
 $DZ(p,n)$, &
 $\{1,5,9,\dots,4\left[\frac{p}8\right]+1\}$ &
 \\
  $p/4\ge 2$   & $\cup
 \{2k+1|k\in \mathbb{N},k\ge 2\left[\frac{p}8\right]+1\}$               &         $\Big( \big[\frac{2(\beta-1)}p \big]+2 \left[\frac{\beta-1}4\right] - \frac{\beta-3}2 \Big)\Big( 2\,\big[\frac{\beta-1}{2\,n}\big]+1\Big)$  \\
$\hbox{gcd}(p,n)=1$  &               &           \\
 \hline
     &               &           \\
 $TZ(24,n)$ & $\{1,7,9\}$ &
 $\Big( 2 \left[\frac{\beta-1}6\right] + \left[\frac{\beta-1}4\right] -
 \frac{\beta-3}2\Big)\Big( 2\,\big[\frac{\beta-1}{2\,n}\big]+1 \Big)$ \\
$\hbox{gcd}(24,n)=1$     &   $\cup \{2k+1|k\in \mathbb{N},k\ge 6\}$             &           \\
 \hline
     &               &           \\
 $OZ(48,n)$ & $\{1,9,13,17,19,21\}$ &
 $\Big( \left[\frac{\beta-1}8\right] + \left[\frac{\beta-1}6\right] +
 \left[\frac{\beta-1}4\right] - \frac{\beta-3}2\Big)\Big( 2\,\big[\frac{\beta-1}{2\,n}\big]+1 \Big)$ \\
$\hbox{gcd}(48,n)=1$      &    $\cup \{2k+1|k\in \mathbb{N}, k\ge 12\}$           &           \\
 \hline
     &   $\{1,13,21,25,31,33,37\}$            &           \\
 $IZ(120,n)$ & $\cup \{41,43,45,49,51,53,55,57\}$ &
 $\Big( \left[\frac{\beta-1}{10}\right] + \left[\frac{\beta-1}6\right] +
 \left[\frac{\beta-1}4\right] - \frac{\beta-3}2 \Big)\Big( 2\,\big[\frac{\beta-1}{2\,n}\big]+1 \Big)$ \\
 $\hbox{gcd}(120,n)=1$& $\cup \, \{2k+1| k\in \mathbb{N}, k \ge 30\}$ & \\
 \hline
 \end{tabular}
 \caption{\label{Tab:Spectrum}
The spectrum of the eigenvalues $E_{\beta}=\beta^2-1$ 
of the Laplace-Beltrami operator on double-action manifolds ${\cal M}$
and their multiplicities $r^{\cal M}(\beta)$ are given \cite{Lustig_2007}.
The bracket $[x]$ denotes the integer part of $x$.
}
\end{table}

$DZ(p,n)$ are inhomogeneous manifolds.
For this reason the eigenmodes on the manifold $DZ(p,n)$ 
depend on the transformation to a new observer 
as discussed in section \ref{sec:obsv_trafo}. 
The corresponding operator can be given by 
\begin{equation}
\label{Eq:operator_new_obs}
D(t) \; = \;
D(\alpha+\epsilon,2\rho,\alpha-\epsilon) \; = \;
e^{\hbox{\scriptsize i}(\alpha+\epsilon) J_{bz}} \,
e^{\hbox{\scriptsize i}(2\rho) J_{by}} \,
e^{\hbox{\scriptsize i}(\alpha-\epsilon)  J_{bz}}
\hspace{10pt},
 \end{equation}
where the coordinates (\ref{Eq:coordinate_t_rho_alpha_epsilon}) are used for 
the observer.
For the following applications it is convenient to transform 
the eigenmodes $| j; m_a, m_b \rangle$ into the spherical basis
$| j; l, m \rangle$, 
where $l$ is the eigenvalue of $\vec L := \vec J_a + \vec J_b$.
These two sets of eigenmodes are connected by 
\begin{eqnarray}
\label{Eq:Trafo_Clebsch_Gordan}
| j; m_a, m_b \rangle & = & \sum_l \langle j m_a j m_b | lm \rangle
\; | j; l, m \rangle
\hspace{10pt} ,  \\
\nonumber
| j; l, m \rangle & = & \sum_{m_a} \langle j m_a j m_b | lm \rangle
\; | j; m_a, m_b \rangle
\hspace{10pt} ,
\end{eqnarray}
where the $\langle j m_a j m_b | lm \rangle$ are the
Clebsch-Gordan coefficients \cite{Edmonds_1957}.
In general, $\langle j m_a j m_b | lm \rangle\neq 0$ only for $0\leq l\leq 2j$
and $m_a+m_b=m$.
Therefore, the expansion with respect to the spherical basis 
$| j; l, m \rangle$ of the eigenmodes $|j,i \rangle$,
eq.\,(\ref{Eq:eigenmodes_DZ}), on $DZ(p,n)$ 
for an arbitrary observer results in
\begin{eqnarray}
\label{Eq:eigenfunction_DZ_exp_sph}
D(t^{-1})|j,i \rangle
\nonumber =
\sum_{l=0}^{2j}\sum_{m=-l}^{l} \xi^{j,m_a(i),m_b(i)}_{lm}(DZ(p,n);t)\,
|j; l, m \rangle
\hspace{10pt} , \\
\nonumber
\xi^{j,m_a,m_b}_{lm}(DZ(p,n);t)\\
=\left\{ 
\begin{array}{ll}\langle j0jm|lm\rangle \,D^{\,j}_{m,m_b}(t^{-1}) 
&: \; j \;\hbox{even}, m_a=0 \\
\frac{1}{\sqrt{2}}\Big(\langle jm_ajm-m_a|lm\rangle \,D^{\,j}_{m-m_a,m_b}(t^{-1}) &\\
\hspace{16pt}+ (-1)^{j+m_a}\langle jm_ajm+m_a|lm\rangle \,D^{\,j}_{m+m_a,m_b}(t^{-1})\Big) 
&: m_a>0\ 
\end{array}\right.\\
\nonumber \hbox{with}\hspace{10pt}
 m_a\equiv 0\;\hbox{mod}\;p/4 \hspace{10pt}\hbox{and}\hspace{10pt}2\,m_b\equiv 0\;\hbox{mod}\;n
\hspace{10pt}. 
\end{eqnarray}
Here the definition of the Wigner polynomial
\begin{equation}
\label{Eq:D_function_rho_alpha_epsilon}
D^{\,j}_{\tilde{m}_b,m_b}(t)
\; := \; \langle j, \tilde{m}_b |D(t)| j,  m_b \rangle
\; = \; 
e^{\hbox{i}\,(\alpha + \epsilon)\,\tilde{m}_b}
d^{\,j}_{\tilde{m}_b, m_b}(2 \rho)
e^{\hbox{i}\,(\alpha - \epsilon)\, m_b}
\end{equation}
is used. 

The calculation of the ensemble average of
the temperature 2-point correlation function $C(\vartheta)$
or the multipole spectrum $C_l$ on the manifolds $DZ(p,n)$ demands 
the computation of the quadratic sum of the expansion coefficients
$\xi^{j,m_a(i),m_b(i)}_{lm}(DZ(p,n);t)$.
This quadratic sum can be rewritten as
\begin{eqnarray}
\nonumber
\frac{1}{2l+1}\sum_{m=-l}^{l}& &\sum_{i=1}^{r^{\cal M}(\beta)}
\left|\xi^{j,m_a(i),m_b(i)}_{lm}(DZ(p,n);t)\right|^2 \\
\label{Eq:quadratic_sum_DZ}
=\frac{1}{2l+1}\sum_{m=-l}^{l}
& &\Big\{{\sum_{m_a,m_b}}'\Big[\langle jm_ajm-m_a|lm\rangle\,
d^{\,j}_{m-m_a,m_b}(-2 \rho)\Big]^2\\
\nonumber& & +{\sum_{m_a>0,m_b}}'\hspace*{2pt}
\Big[(-1)^{j+m_a}\langle jm_ajm-m_a|lm\rangle\,\langle jm_ajm+m_a|lm\rangle \\
\nonumber& & 
\hspace{32pt}d^{\,j}_{m-m_a,m_b}(-2 \rho)\,d^{\,j}_{m+m_a,m_b}(-2 \rho)
\cos(2\,m_a(\alpha-\epsilon))\Big]
\Big\}
\hspace{10pt} .
\end{eqnarray}
The primes at the sums indicate
that the summation is restricted by the conditions
$m_a\equiv 0 \hbox{ mod }p/4$ and $2\,m_b\equiv 0 \hbox{ mod }n$.
Therefore, the analysis of the CMB statistics can be confined 
to observer positions within the $\rho$-$\alpha$ plane by setting $\epsilon=0$.
Furthermore, taking into account the condition $m_a\equiv 0\;\hbox{mod}\;p/4$
and the symmetry of the sum (\ref{Eq:quadratic_sum_DZ}) with respect to the
transformation $\rho \rightarrow \frac{\pi}{2}-\rho$,
the domain of observer positions exhausting the complete CMB variability
can be reduced to the smaller intervals
$\alpha \in [0,\frac{\pi}{p}]$ and $\rho \in [0,\frac{\pi}{4}]$.

\section{CMB Anisotropy on Large Angular Scales}
\label{sec:S_statistics}

The quadratic sum (\ref{Eq:quadratic_sum_DZ}) of the expansion coefficients
$\xi^{j,i}_{lm}:=\xi^{j,m_a(i),m_b(i)}_{lm}$ allows the computation of
the ensemble average of the multipole moments $C_l$ for the
space ${\cal M}=DZ(p,n)$
\begin{eqnarray}
C_l & := & \nonumber
\frac{1}{2l+1}\sum_{m=-l}^l\left\langle\left|a_{lm}\right|^2\right\rangle
\\ & = &
\label{Eq:Cl_ensemble}
\sum_{\beta}\frac{ T_l^2(\beta) \; P(\beta)}{2l+1}\sum_{m=-l}^{l}
\sum_{i=1}^{r^{\cal M}(\beta)} \left|\xi^{\beta,i}_{lm}({\cal M};t)\right|^2
\hspace{10pt} .
\end{eqnarray}
The initial power spectrum is
$P(\beta)\sim 1/(E_{\beta}\,\beta^{2-n_{\hbox{\scriptsize s}}})$
and $T_l^2(\beta)$ is the transfer function
for which the same cosmological model as in \cite{Aurich_Lustig_2012a}
is used.
After the multipole moments $C_l$ have been obtained,
the ensemble average of the 2-point correlation function $C(\vartheta)$
can be computed using
\begin{equation}
\label{Eq:C_theta_Cl}
C(\vartheta) \; = \; 
\sum_l\,\frac{2l+1}{4\pi}\,C_l\,P_l\left(\cos\vartheta\right)
\hspace{10pt}.
\end{equation}
This in turn leads to the $S$ statistics using eq.\,(\ref{Eq:S_statistic_60}),
and the extent of the suppression of the CMB correlations on
large angular scales can be calculated.
In the following, we always normalise the $S$ statistics to that of the
homogeneous ${\cal S}^3$ space.
Thus, values of the $S$ statistics below one indicate models
that possess a stronger CMB suppression than the model based on the
simply connected spherical ${\cal S}^3$ space.

We compute the $S$ statistics along these lines for
sets of cosmological parameters which are close to the
standard concordance model of cosmology \cite{Larson_et_al_2011}.
The parameters are obtained from the LAMBDA website (lambda.gsfc.nasa.gov),
see the WMAP cosmological parameters of the model 'olcdm+sz+lens'
using the data 'wmap7+bao+snconst',
which are $\Omega_{\hbox{\scriptsize b}} = 0.0485$,
$\Omega_{\hbox{\scriptsize cdm}} = 0.238$, the Hubble constant $h=0.681$,
and the spectral index $n_{\hbox{\scriptsize s}}=0.961$.
The density parameter of the cosmological constant 
$\Omega_{\scriptsize \Lambda }$ is varied such that the total 
density parameter $\Omega_{\hbox{\scriptsize tot}}$ is
in the interval $\Omega_{\hbox{\scriptsize tot}}=1.001,\dots,1.05$.
Thus, we consider spherical models that are almost flat.
The cosmological parameters stated above lead to the constraint
$0.99 < \Omega_{\hbox{\scriptsize tot}} < 1.02$ (95\% CL),
so that our chosen $\Omega_{\hbox{\scriptsize tot}}$ interval covers slightly
more than 99\% CL.
For each value of $\Omega_{\hbox{\scriptsize tot}}$
the correlation measure $S$ is calculated for numerous observer positions
described by the parameters $\alpha$ and $\rho$.
The values of $\alpha$ and $\rho$ are obtained from a sufficiently dense
rectangular mesh with $\alpha \in [0,\frac{\pi}{p}]$ and
$\rho \in [0,\frac{\pi}{4}]$.
This leads to almost 4.3 million simulations
up to $\Omega_{\hbox{\scriptsize tot}}= 1.05$.


\begin{figure}
\begin{center}
\begin{minipage}{10cm}
\vspace*{-25pt}
\hspace*{18pt}\includegraphics[width=10.0cm]{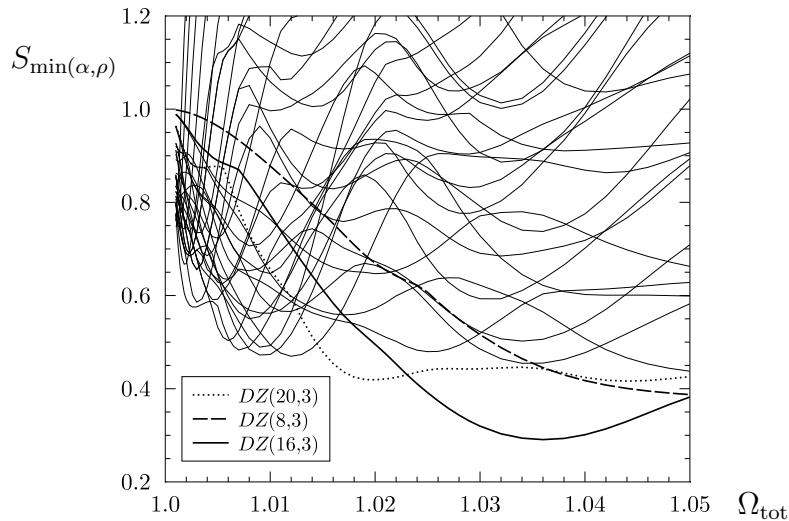}
\put(-25,26){$\Omega_{\hbox{\scriptsize tot}}$}
\put(-300,195){$S_{\hbox{\scriptsize min}(\alpha,\rho)}$}
\end{minipage}
\vspace*{-25pt}
\end{center}
\caption{\label{Fig:S60_Otot_1.05}
The minima of the $S$ statistics, eq.\,(\ref{Eq:S_statistic_60}),
taken over all positions in the $\alpha$-$\rho$ plane,
are plotted as a function of $\Omega_{\hbox{\scriptsize tot}}$
for all $DZ(p,n)$ spaces with $|\Gamma|=p\,n\leq 180$.
The correlation measure $S$ is normalised to that of the ${\cal S}^3$ space.
}
\end{figure}


The minima of the $S$ statistics
\begin{equation}
\label{Eq:S_min}
S_{\hbox{\scriptsize min}(\alpha,\rho)}=\hbox{min}_{\{ \alpha,\rho \}}\left(\frac{S( \alpha,\rho)}{S_{{\cal S}^3}}\right)
\end{equation}
are determined for fixed values of $\Omega_{\hbox{\scriptsize tot}}$
for all 33 topologies $DZ(p,n)$ with a group order up to 180
and are plotted in figure \ref{Fig:S60_Otot_1.05}.
There are several spaces that have a CMB suppression
more than two times stronger than in the ${\cal S}^3$ space.
The $DZ(16,3)$ space has a minimum at $\Omega_{\hbox{\scriptsize tot}}=1.036$
with $S_{\hbox{\scriptsize min}(\alpha,\rho)}=0.291$.
Thus it has three times smaller correlations compared to
the ${\cal S}^3$ space.
Notice that this model has at the upper boundary of the 95\% CL interval,
i.\,e.\ at $\Omega_{\hbox{\scriptsize tot}}=1.02$,
a noteworthy suppression factor of about 0.5.
Another candidate is provided by the double-action space $DZ(20,3)$
which has two almost equal minima at $\Omega_{\hbox{\scriptsize tot}}=1.020$
and at $\Omega_{\hbox{\scriptsize tot}}=1.044$.
The minima are $S_{\hbox{\scriptsize min}(\alpha,\rho)}=0.419$
and $S_{\hbox{\scriptsize min}(\alpha,\rho)}=0.416$, respectively.
Therefore, if one allows for $\Omega_{\hbox{\scriptsize tot}}$ only values
as large as $\Omega_{\hbox{\scriptsize tot}}=1.02$,
the best candidate would be provided by the $DZ(20,3)$ space.
If one restricts $\Omega_{\hbox{\scriptsize tot}}$ to even smaller values,
that is, to even flatter models,
then several other $DZ(p,n)$ spaces possess the lowest CMB correlations.
With $\Omega_{\hbox{\scriptsize tot}}<1.015$
one finds the four models $DZ(24,5)$, $DZ(24,7)$, $DZ(28,5)$,
and $DZ(32,5)$, satisfying $S_{\hbox{\scriptsize min}(\alpha,\rho)}<0.5$.
The corresponding values of $S_{\hbox{\scriptsize min}(\alpha,\rho)}$
are given in table \ref{Tab:double_action},
where the parameters of the best models are listed for all 33 manifolds
restricted to $\Omega_{\hbox{\scriptsize tot}}\leq 1.05$.


\begin{table}[htbp]
\centering
\begin{tabular}{|c|c|c|c|c|}
\hline
manifold ${\cal M}$ &
$S_{\hbox{\scriptsize min}(\Omega_{\hbox{\scriptsize tot}},\alpha,\rho)}$ &
$\Omega_{\hbox{\scriptsize tot}}$  & $\rho$ &  $\alpha$ \\
\hline
$DZ(8,3)$ & 0.387 & 1.050 & 0.479 & 0.785 \\
$DZ(8,5)$ & 0.647 & 1.020 & 0.393 & 0.000 \\
$DZ(8,7)$ & 0.707 & 1.008 & 0.212 & 0.000 \\
$DZ(8,9)$ & 0.723 & 1.005 & 0.620 & 0.060 \\
$DZ(8,11)$ & 0.735 & 1.003 & 0.668 & 0.112 \\
$DZ(8,13)$ & 0.737 & 1.002 & 0.770 & 0.224 \\
$DZ(8,15)$ & 0.743 & 1.002 & 0.691 & 0.071 \\
$DZ(8,17)$ & 0.745 & 1.001 & 0.738 & 0.150 \\
$DZ(8,19)$ & 0.748 & 1.001 & 0.738 & 0.117 \\
$DZ(8,21)$ & 0.760 & 1.001 & 0.738 & 0.117 \\
$DZ(12,5)$ & 0.437 & 1.050 & 0.385 & 0.000 \\
$DZ(12,7)$ & 0.599 & 1.050 & 0.369 & 0.000 \\
$DZ(12,11)$ & 0.656 & 1.003 & 0.055 & 0.000 \\
$DZ(12,13)$ & 0.690 & 1.002 & 0.047 & 0.349 \\
$DZ(16,3)$ & 0.291 & 1.036 & 0.385 & 0.000 \\
$DZ(16,5)$ & 0.480 & 1.025 & 0.408 & 0.393 \\
$DZ(16,7)$ & 0.550 & 1.008 & 0.181 & 0.178 \\
$DZ(16,9)$ & 0.626 & 1.006 & 0.086 & 0.000 \\
$DZ(16,11)$ & 0.586 & 1.003 & 0.055 & 0.262 \\
$DZ(20,3)$ & 0.416 & 1.044 & 0.393 & 0.314 \\
$DZ(20,7)$ & 0.562 & 1.010 & 0.605 & 0.000 \\
$DZ(20,9)$ & 0.591 & 1.005 & 0.149 & 0.035 \\
$DZ(24,5)$ & 0.470 & 1.012 & 0.565 & 0.262 \\
$DZ(24,7)$ & 0.489 & 1.009 & 0.605 & 0.262 \\
$DZ(28,3)$ & 0.455 & 1.034 & 0.385 & 0.224 \\
$DZ(28,5)$ & 0.472 & 1.009 & 0.589 & 0.000 \\
$DZ(32,3)$ & 0.502 & 1.034 & 0.377 & 0.172 \\
$DZ(32,5)$ & 0.483 & 1.007 & 0.613 & 0.196 \\
$DZ(36,5)$ & 0.503 & 1.006 & 0.620 & 0.000 \\
$DZ(40,3)$ & 0.685 & 1.032 & 0.385 & 0.157 \\
$DZ(44,3)$ & 0.656 & 1.003 & 0.055 & 0.000 \\
$DZ(52,3)$ & 0.690 & 1.002 & 0.047 & 0.121 \\
$DZ(56,3)$ & 0.667 & 1.002 & 0.039 & 0.112 \\
\hline
\end{tabular}
\caption{\label{Tab:double_action}
For the 33 prism double-action manifolds
$DZ(p,n) = {\cal S}^3/\left(D^\star_p \times Z_n\right)$
up to the group order $|\Gamma|=180$,
the models with the lowest CMB correlations on large angular scales
are given.
The parameters $\Omega_{\hbox{\scriptsize tot}}$, $\rho$, and $\alpha$
specify the model which leads to a minimal value in the $S$ statistics.
}
\end{table}


The only manifold where the first minimum of
$S_{\hbox{\scriptsize min}(\alpha,\rho)}$
lies outside this $\Omega_{\hbox{\scriptsize tot}}$ interval is $DZ(8,3)$
such that its value in table \ref{Tab:double_action} is determined by
this $\Omega_{\hbox{\scriptsize tot}}$ cut-off.
The minimum with $S_{\hbox{\scriptsize min}(\alpha,\rho)}=0.060$ occurs
at $\Omega_{\hbox{\scriptsize tot}}=1.15$ with $\alpha=\rho=0$.
Although $DZ(8,3)$ has a very strong suppression,
this value of $\Omega_{\hbox{\scriptsize tot}}$ is too large
in order to be compatible with the current cosmological observations.


\begin{figure}
\begin{center}
\begin{minipage}{10cm}
\vspace*{-25pt}
\hspace*{18pt}\includegraphics[width=10.0cm]{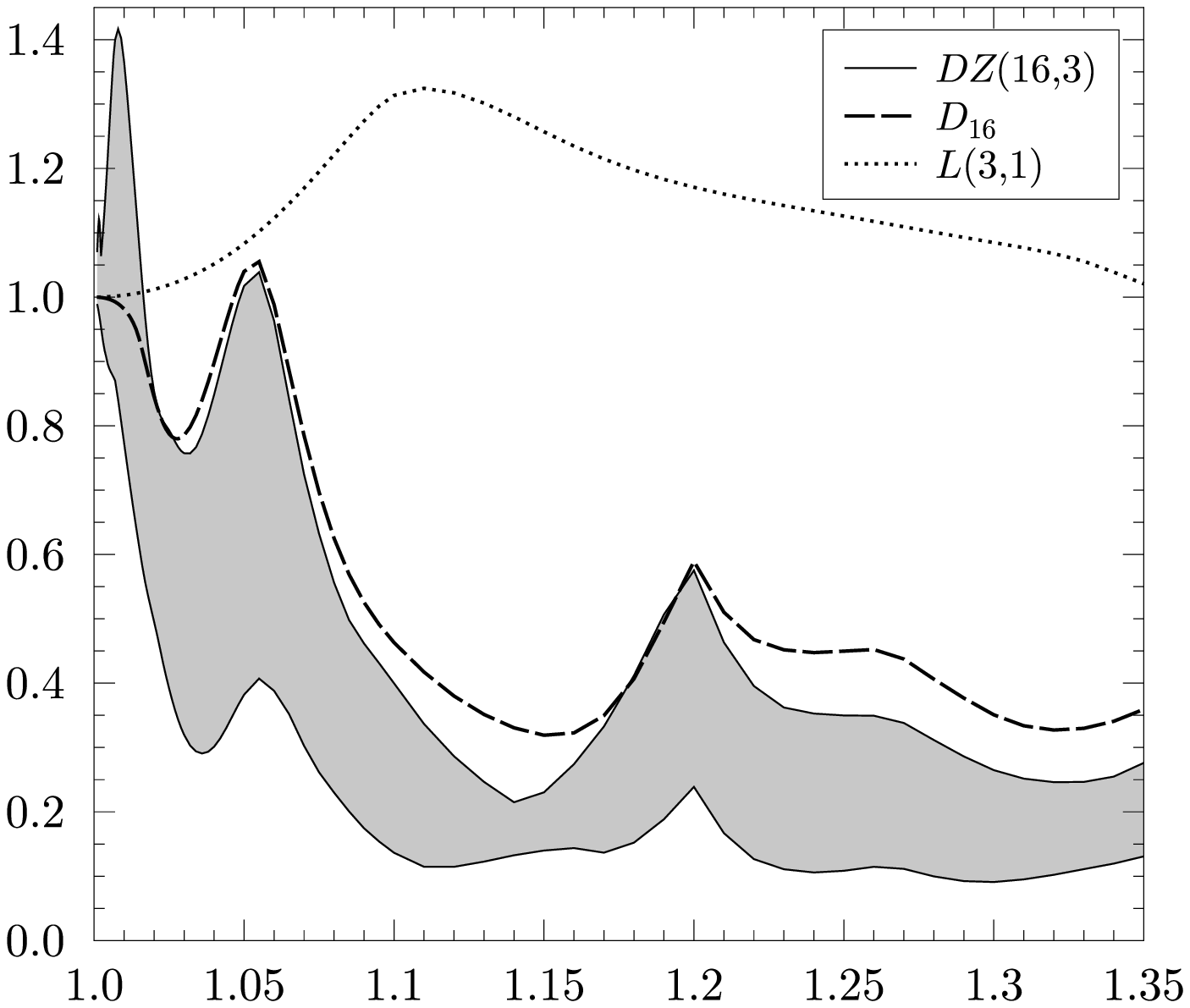}
\put(-215,195){(a)}
\put(-25,26){$\Omega_{\hbox{\scriptsize tot}}$}
\put(-275,195){$\frac{S}{S_{{\cal S}^3}}$}
\end{minipage}
\begin{minipage}{10cm}
\vspace*{-45pt}
\hspace*{18pt}\includegraphics[width=10.0cm]{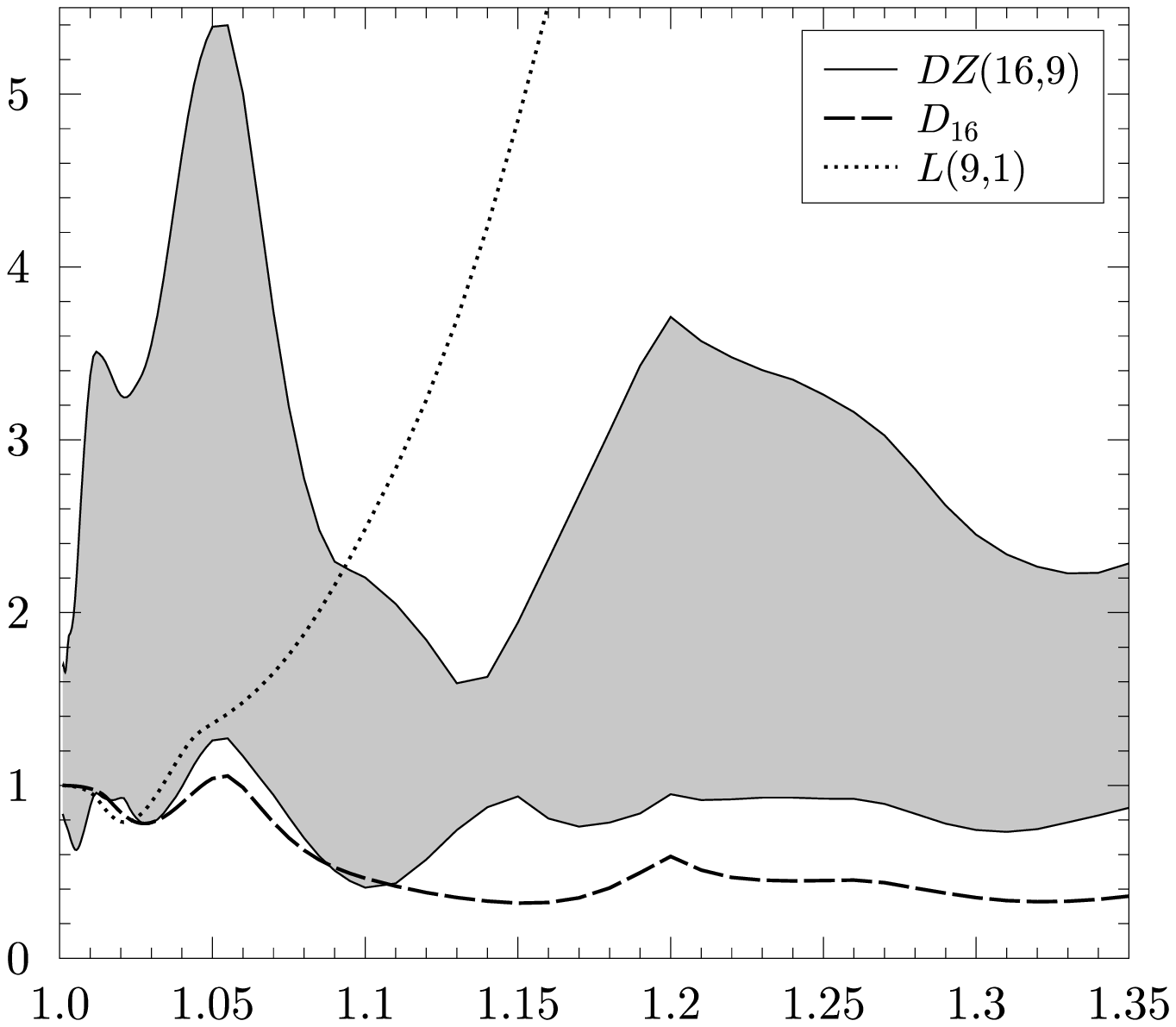}
\put(-200,195){(b)}
\put(-25,26){$\Omega_{\hbox{\scriptsize tot}}$}
\put(-275,195){$\frac{S}{S_{{\cal S}^3}}$}
\end{minipage}
\vspace*{-25pt}
\end{center}
\caption{\label{Fig:S60_variation_DZ_16_3}
The panel (a) compares the $S$ statistics of the homogeneous prism space
$D_{16}$ and the homogeneous lens space $L(3,1)$
with that of the inhomogeneous double-action space $DZ(16,3)$.
For the latter the variation of the $S$ statistics due to the position
dependence is shown as a grey band.
It is obvious that the $DZ(16,3)$ space inherits its CMB properties
from the $D_{16}$ space but not from the lens space $L(3,1)$.
The corresponding comparison of $DZ(16,9)$ with
$D_{16}$ and $L(9,1)$ is shown in panel (b).
Note that both panels use a different scale and that the
variation of the $S$ statistics is much larger in the case
$DZ(16,9)$.
}
\end{figure}



\begin{figure}
\begin{center}
\vspace*{-12pt}
\hspace*{40pt}\includegraphics[width=10.0cm]{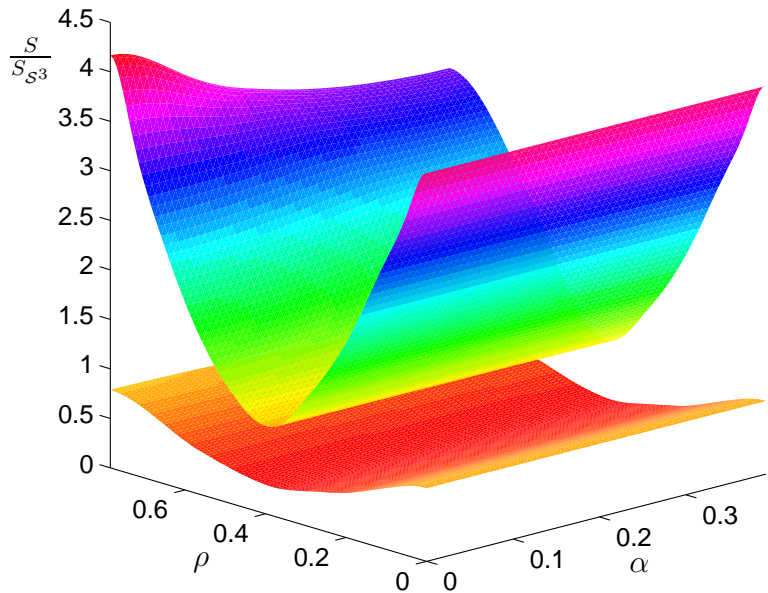}
\put(-300,205){$\frac{S}{S_{{\cal S}^3}}$}
\put(-230,15){$\rho$}
\put(-65,13){$\alpha$}
\end{center}
\caption{\label{Fig:S60_DZ_16_3_DZ_16_9_rho_alpha}
The normalised correlation measure $S$ is shown in dependence on the observer
position which is parameterised by $\rho$ and $\alpha$.
The lower surface belongs to the double-action space $DZ(16,3)$
at $\Omega_{\hbox{\scriptsize tot}}=1.036$.
It has only a mild position dependence compared to the manifold $DZ(16,9)$
(upper surface) computed for the same $\Omega_{\hbox{\scriptsize tot}}$.
The figure reveals that the position dependence is more pronounced
with respect to the parameter $\rho$ than to $\alpha$ in both cases.
}
\end{figure}



\begin{table}[htbp]
\centering
\begin{tabular}{|c|c|c|}
\hline
manifold ${\cal M}$ &
$S_{\hbox{\scriptsize min}(\Omega_{\hbox{\scriptsize tot}})}$ &
$\Omega_{\hbox{\scriptsize tot}}$ \\
\hline
$D_{8}$ & 0.988 & 1.050 \\ 
$D_{12}$ & 0.684 & 1.050 \\ 
$D_{16}$ & 0.780 & 1.028 \\ 
$D_{20}$ & 0.806 & 1.017 \\ 
$D_{24}$ & 0.822 & 1.012 \\ 
$D_{28}$ & 0.831 & 1.008 \\ 
$D_{32}$ & 0.838 & 1.007 \\ 
$D_{36}$ & 0.842 & 1.005 \\ 
$D_{40}$ & 0.846 & 1.004 \\ 
$D_{44}$ & 0.852 & 1.004 \\ 
$D_{48}$ & 0.854 & 1.003 \\ 
$D_{52}$ & 0.854 & 1.002 \\ 
$D_{56}$ & 0.855 & 1.002 \\ 
$D_{60}$ & 0.857 & 1.002 \\ 
$D_{64}$ & 0.858 & 1.002 \\ 
$D_{68}$ & 0.859 & 1.001 \\ 
$D_{72}$ & 0.860 & 1.001 \\ 
\hline
\end{tabular}
\caption{\label{Tab:dieder}
The lowest CMB correlations on large angular scales
$S_{\hbox{\scriptsize min}(\Omega_{\hbox{\scriptsize tot}})}$
are given with the corresponding $\Omega_{\hbox{\scriptsize tot}}$
for all prism spaces $D_p$ generated by the binary dihedral groups $D^\star_p$
up to the group order $|\Gamma|=p=72$.
The minima are determined for the interval
$\Omega_{\hbox{\scriptsize tot}}=1.001\dots 1.05$.
The prism spaces are homogeneous, and thus
an observer position is not required in this table.
}
\end{table}


Let us discuss the favourite $DZ(16,3)$ in more detail.
Since figure \ref{Fig:S60_Otot_1.05} only shows the minimum
$S_{\hbox{\scriptsize min}(\alpha,\rho)}$,
the degree of variation with respect to the observer position
parameterised by $\alpha$ and $\rho$ is eliminated.
This information is provided in figure \ref{Fig:S60_variation_DZ_16_3}(a)
where the variation due to the position is shown as a grey band
for the $DZ(16,3)$ space.
In addition, the panel shows the normalised correlation measure $S$ of
the homogeneous $D_{16}$ and $L(3,1)$ spaces (dashed and dotted curves)
which are generated by the groups $D^\star_{16}$ and $Z_3$.
Since $DZ(16,3) = {\cal S}^3/\left(D^\star_{16} \times Z_3\right)$,
these are the subgroups $R$ and $L$ generating $DZ(16,3)$.
The figure \ref{Fig:S60_variation_DZ_16_3}(a) reveals
that it is the behaviour of the homogeneous $D_{16}$ space
which is responsible for the main behaviour of the inhomogeneous $DZ(16,3)$.
Because of this relevance for the $DZ(p,n)$ spaces,
the table \ref{Tab:dieder} lists the minima of their $S$ statistics
together with the value of $\Omega_{\hbox{\scriptsize tot}}$
where the minima occur.
As in previous cases the $\Omega_{\hbox{\scriptsize tot}}$ interval is
restricted to $\Omega_{\hbox{\scriptsize tot}}=1.001\dots 1.05$.
The table reveals that all prism spaces $D_p$, $p\leq 72$,
possess only a moderate suppression of large-angle correlations below
$\Omega_{\hbox{\scriptsize tot}}=1.05$.
The minimum for $D_{16}$ at $\Omega_{\hbox{\scriptsize tot}}=1.028$
can also be seen in figure \ref{Fig:S60_variation_DZ_16_3} (dashed curves).

There are five manifolds $DZ(16,n)$ up to group order 180
which have the group $D^\star_{16}$ as a subgroup.
As a further example the variation due to the observer position
in $DZ(16,9)$ is shown in figure \ref{Fig:S60_variation_DZ_16_3}(b)
where the variation is larger than for the manifold $DZ(16,3)$.
The increased variability can be understood in terms of the number of
inhomogeneous translations within the deck group.
An inhomogeneous group element transforms different points
$\vec x \in {\cal S}^3$ to varying spherical distances.
This contrasts to Clifford transformations where all points
$\vec x \in {\cal S}^3$ are shifted by the same spherical distance.
The deck group of the manifold $DZ(16,3)$ contains 30 inhomogeneous
transformations and 18 Clifford transformations.
The number of inhomogeneous translations increases to 120 for 
the manifold $DZ(16,9)$.
The large variability of the $S$ statistics can be explained
by this large number,
since the CMB dependence on the observer position is the more pronounced,
the more inhomogeneous translations are in the deck group.
Conversely, the variation width must shrink to zero if all group elements
are Clifford transformations whose transformation properties are
independent of $\vec x \in {\cal S}^3$.
To emphasise this point
figure \ref{Fig:S60_DZ_16_3_DZ_16_9_rho_alpha} displays the
observer position dependence for the manifolds $DZ(16,3)$ and $DZ(16,9)$
of the normalised $S$ statistics computed at the same value of
$\Omega_{\hbox{\scriptsize tot}}=1.036$.
It is clearly seen that the $DZ(16,9)$ space has a severe position
dependence compared to the space $DZ(16,3)$.
Furthermore, both models posses only a modest variation with respect
the the parameter $\alpha$.
Thus, the main variation is due to an observer shift in the $\rho$ direction.

\section{The shape of the Dirichlet domains}
\label{sec:Dirichlet}


\begin{figure}
\vspace*{-180pt}\hspace*{35pt}\includegraphics[width=15.0cm]{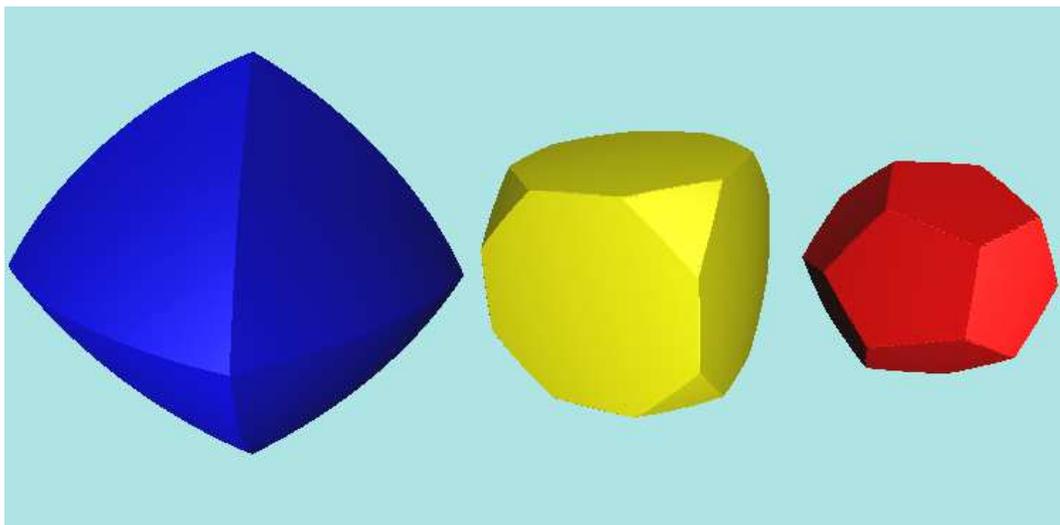}
\vspace*{-215pt}\caption{\label{Fig:Dirichlet_Polyeder}
The Dirichlet domains generated by
the binary tetrahedral group $T^\star$ (blue),
the binary octahedral group $O^\star$ (yellow),
and the binary icosahedral group $I^\star$ (red) are shown.
These three regular polyhedral spaces are homogeneous manifolds.
}
\end{figure}


As described in section \ref{sec:obsv_trafo}
the representation of the group elements of the deck group $\Gamma$
changes according to eq.\,(\ref{Eq:trafo_group}).
As a consequence the Dirichlet domain alters due to shifts
of the observer position.
The Dirichlet domain ${\cal F}$ is defined as the set of points
$u \in {\cal S}^3$
that cannot be transformed any closer to the observer $u_o$
by applying the elements of the deck group $\Gamma$, i.\,e.\
\begin{equation}
u \in {\cal F} \hspace{5pt} \hbox{ if } \hspace{5pt}
d(u_o, u) \; \leq \; d(u_o, g(u))
\hspace{10pt} \hbox{ for all } g \in \Gamma
\hspace{10pt} ,
\end{equation}
where $d(u_1,u_2)$ measures the spherical distances between the points
$u_1,u_2\in{\cal S}^3$.
It is instructive to compare the shape of the Dirichlet domains
at the positions of the observer where the values for the $S$ statistics
are extremal.
The figure \ref{Fig:Dirichlet_Polyeder} shows the Dirichlet domains of
the spherical regular polyhedral spaces generated by
the binary tetrahedral group $T^\star$,
the binary octahedral group $O^\star$,
and the binary icosahedral group $I^\star$
which are introduced in section \ref{sec:eigenmode}.
These groups consist only of Clifford transformations,
so that the Dirichlet domains are independent of the observer position,
i.\,e.\ these manifolds are homogeneous.

\begin{figure}
\begin{center}
\vspace*{-80pt}
\begin{minipage}{16cm}
\begin{minipage}{8cm}
\hspace*{20pt}\includegraphics[width=10.0cm]{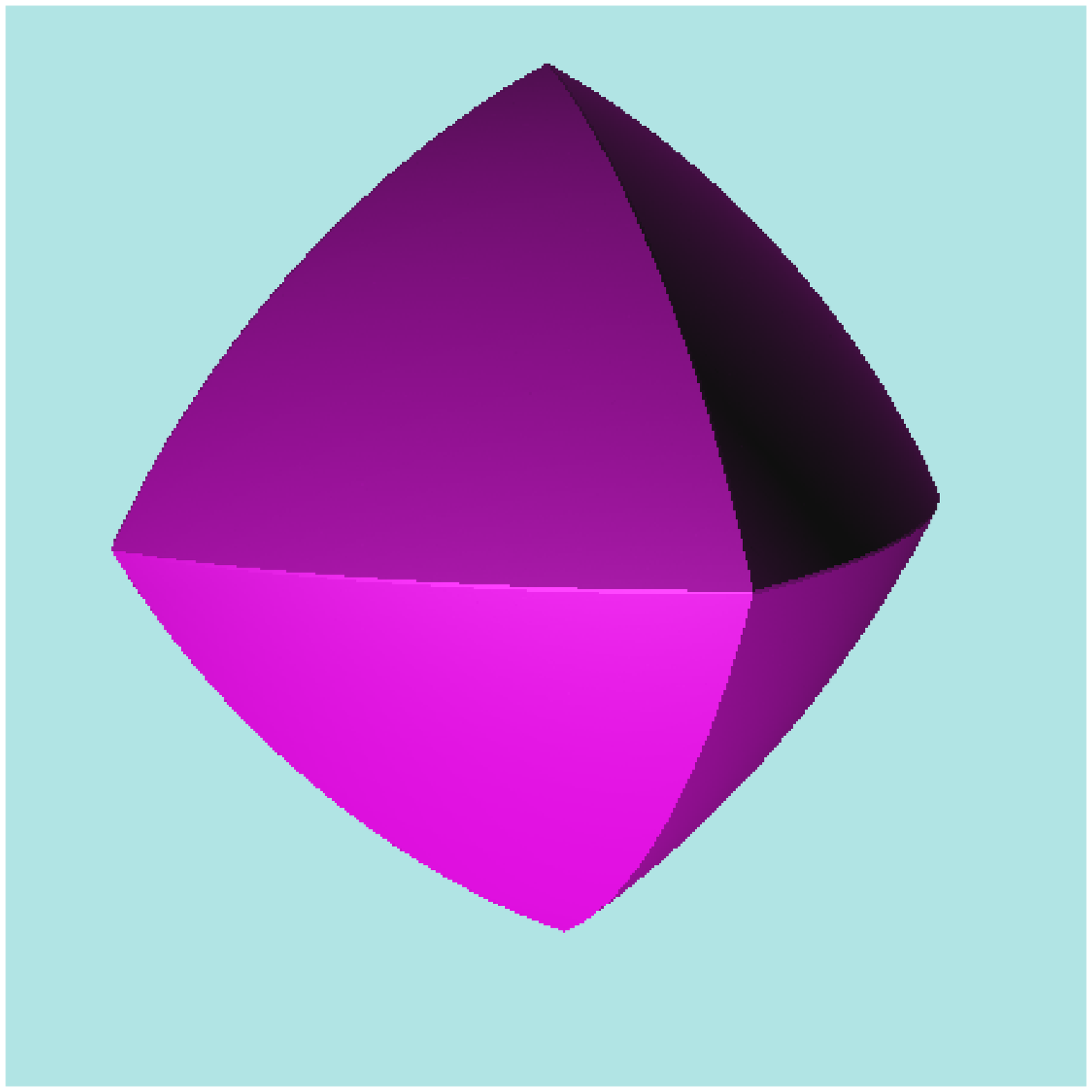}
\put(-220,280){(a)}
\end{minipage}
\begin{minipage}{8cm}
\hspace*{0pt}\includegraphics[width=10.0cm]{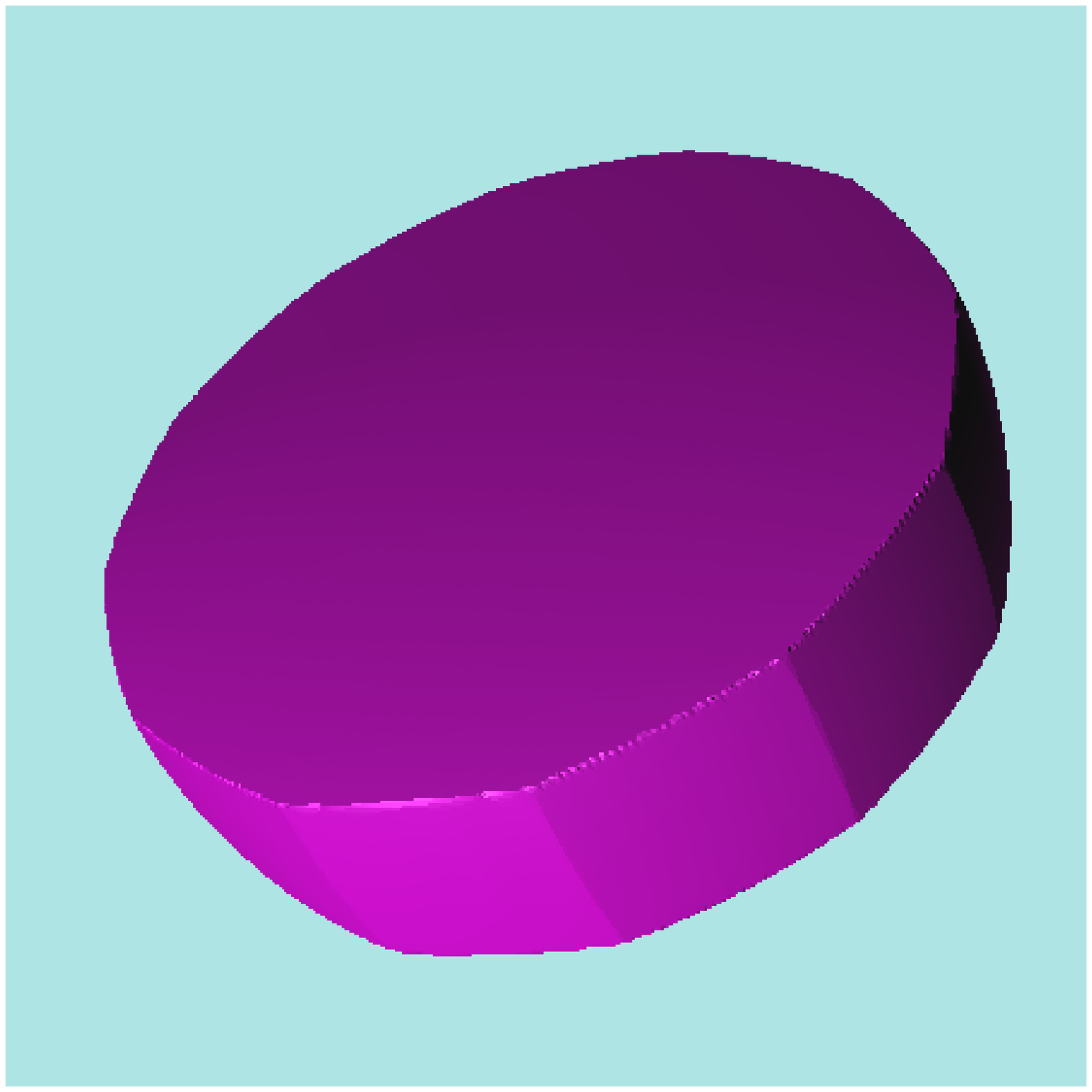}
\put(-220,280){(b)}
\end{minipage}
\end{minipage}
\vspace*{-100pt}
\end{center}
\caption{\label{Fig:Dirichlet_DZ_8_3}
The Dirichlet domains for $DZ(8,3)$ are shown with
$\rho=0.479$ and $\alpha=\frac\pi 4$ in panel (a) and with
$\rho=0$ and $\alpha=0$ in panel (b).
}
\end{figure}

Let us now turn to the $DZ(8,3)$ space which has its first minimum in the
$S$ statistics at $\Omega_{\hbox{\scriptsize tot}}=1.15$.
Figure \ref{Fig:S60_Otot_1.05} shows its behaviour up to
$\Omega_{\hbox{\scriptsize tot}}=1.05$ and the decline towards higher
$\Omega_{\hbox{\scriptsize tot}}$ is already visible.
In the range $\Omega_{\hbox{\scriptsize tot}}=1.025\dots 1.065$
the minimum in the $S$ statistics is achieved at
$\rho=0.479$ and $\alpha=\frac\pi 4$
whereas the maximum occurs at $\rho=0$ and $\alpha=0$.
The Dirichlet domains for these positions are shown in
figure \ref{Fig:Dirichlet_DZ_8_3}.
Interestingly, the orientations are reversed at the actual minimum
at $\Omega_{\hbox{\scriptsize tot}}=1.15$,
i.\,e.\ the minimum occurs at $\rho=0$ and $\alpha=0$
and the maximum at $\rho=0.479$ and $\alpha=\frac\pi 4$.
A further interesting point is that the Dirichlet domain of $DZ(8,3)$
is at $\rho=\frac 12\arccos(1/\sqrt 3)\simeq 0.479$
and $\alpha=\frac\pi 4$ identical to that
of the binary tetrahedral space ${\cal T}$
shown in figure \ref{Fig:Dirichlet_Polyeder}.
It is remarkable
that in the range $\Omega_{\hbox{\scriptsize tot}}=1.025\dots 1.065$
the favoured geometric shape is the same for both the $DZ(8,3)$ space
and the binary tetrahedral space ${\cal T}$,
since the wave number spectrum starts at $\beta=5$ for the former
and at $\beta=7$ for the latter (see table \ref{Tab:Spectrum}).
Furthermore, their multiplicities are also different.
This fact is remarkable in view of the so-called well-proportioned conjecture
\cite{Weeks_Luminet_Riazuelo_Lehoucq_2005}
which states that the CMB suppression on large angular scales is the more
pronounced, the more well-proportioned the Dirichlet domain is.
Thus, one would expect the position at $\rho=0.479$ and $\alpha=\frac\pi 4$
always to be the one with the minimum in the $S$ statistics,
but this is not the case.
This provides therefore a counterexample to the conjecture.
We would like to clarify that although the Dirichlet domains
of the binary tetrahedral space ${\cal T}$ and of the $DZ(8,3)$ space
at $\rho=\frac 12\arccos(1/\sqrt 3)$ and $\alpha=\frac\pi 4$ are identical,
they nevertheless belong to different spaces
since the gluing rules, which describe how to connect the faces,
are different.


\begin{figure}
\begin{center}
\vspace*{-80pt}
\begin{minipage}{16cm}
\begin{minipage}{8cm}
\hspace*{20pt}\includegraphics[width=10.0cm]{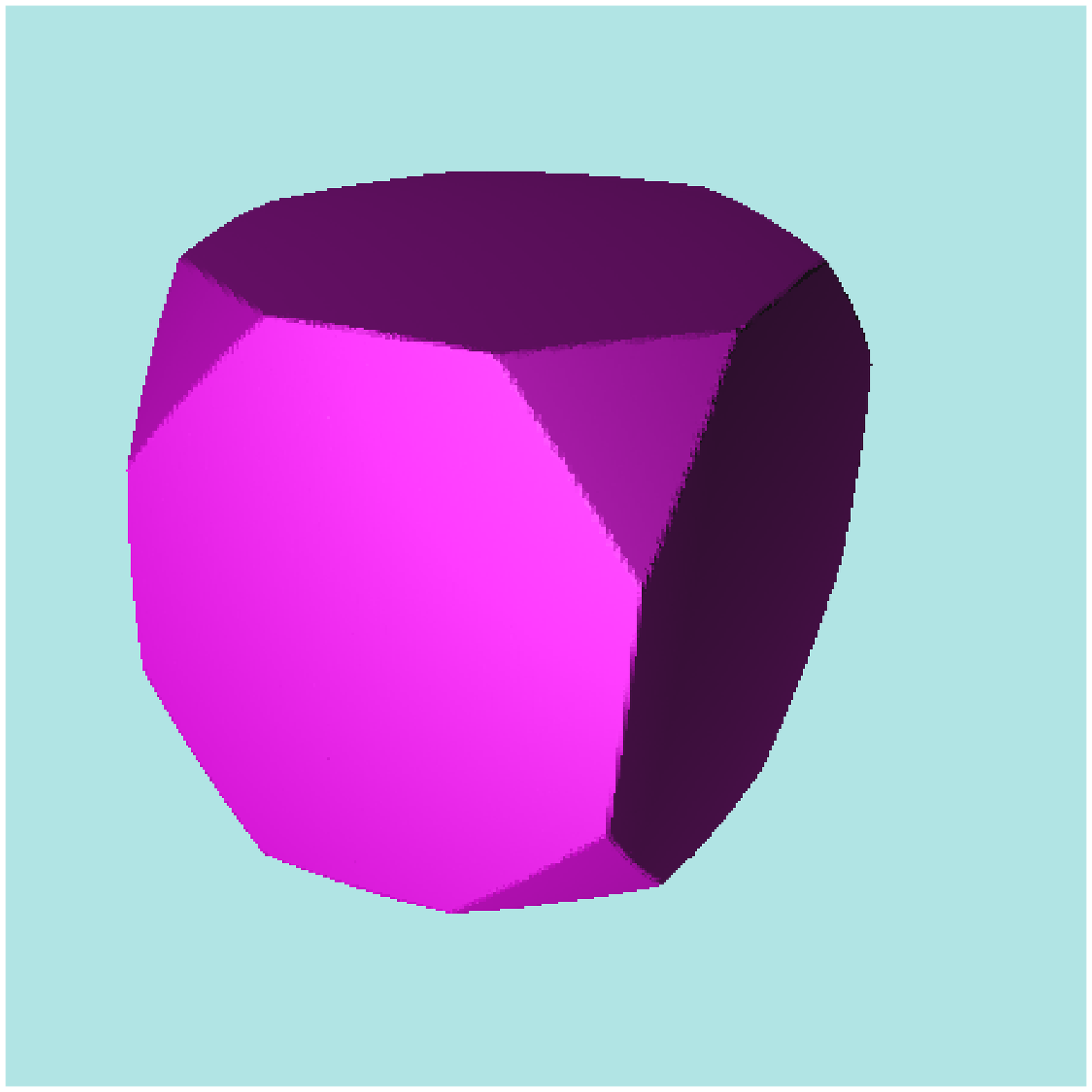}
\put(-220,280){(a)}
\end{minipage}
\begin{minipage}{8cm}
\hspace*{0pt}\includegraphics[width=10.0cm]{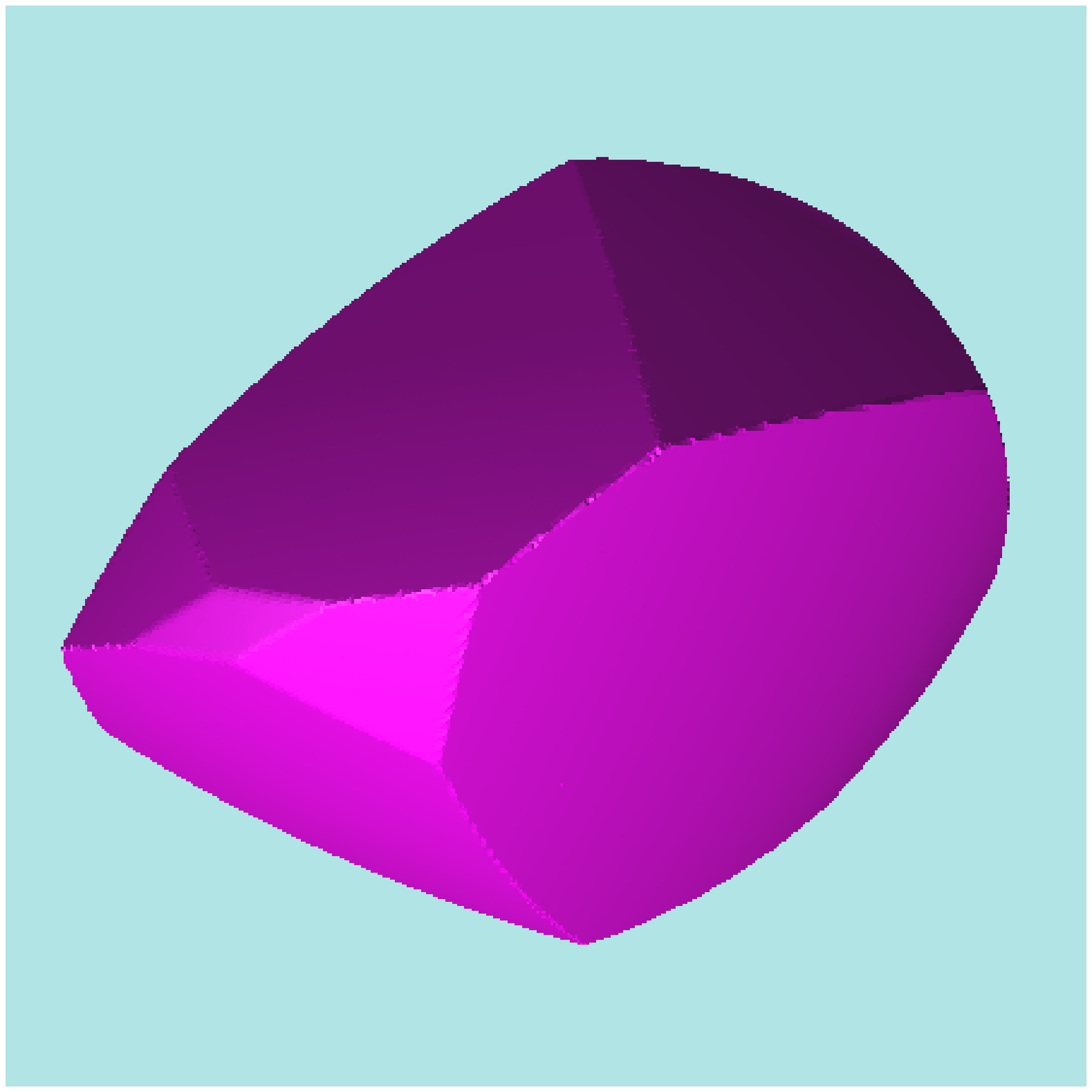}
\put(-220,280){(b)}
\end{minipage}
\end{minipage}
\vspace*{-100pt}
\end{center}
\caption{\label{Fig:Dirichlet_DZ_16_3}
The Dirichlet domains for $DZ(16,3)$ are shown with
$\rho=0.385$ and $\alpha=0$ in panel (a) and with
$\rho=\frac\pi 4$ and $\alpha=0$ in panel (b).
}
\end{figure}

The discussion in section \ref{sec:S_statistics} shows that a very interesting
double-action space is provided by $DZ(16,3)$ having a strong CMB anisotropy
suppression at $\Omega_{\hbox{\scriptsize tot}}=1.036$.
The Dirichlet domain for the observer who sees the minimal CMB anisotropy,
i.\,e.\ $\rho=0.385$ and $\alpha=0$, is depicted
in figure \ref{Fig:Dirichlet_DZ_16_3}(a).
A comparison with figure \ref{Fig:Dirichlet_Polyeder} reveals
the similarity with the homogeneous space generated
by the binary octahedral group $O^\star$.
However, the Dirichlet domain possesses not the symmetry of
the fundamental cell of the group $O^\star$,
since the Dirichlet domain that exactly matches that of
the binary octahedral group
$O^\star$ occurs in $DZ(16,3)$ at $\rho=\frac 12\arccos(1/\sqrt 3)\simeq 0.479$
and $\alpha=0$.
At this position the value for the $S$ statistics is $S=0.346$
which is larger than the minimal value $S=0.291$
as revealed by table \ref{Tab:double_action}.
Therefore, although both Dirichlet domains are similar,
the deviation demonstrates that the best Dirichlet domain
with respect to maximal CMB suppression is not the one with
the most well-proportioned fundamental cell.
This difference is not surprising since the wave number spectrum starts
at $\beta=5$ for $DZ(16,3)$ and at $\beta=9$ for the
binary octahedral space ${\cal O}$ (see table \ref{Tab:Spectrum}).
Figure \ref{Fig:Dirichlet_DZ_16_3}(b) shows the Dirichlet domain of $DZ(16,3)$
at $\rho=\frac\pi 4$ and $\alpha=0$
which corresponds to the observer seeing the largest CMB anisotropy power
on large angular scales at $\Omega_{\hbox{\scriptsize tot}}=1.036$.


The above results reflect the complexity of eq.\,(\ref{Eq:Cl_ensemble})
which is used to compute the CMB correlations.
Eq.\,(\ref{Eq:Cl_ensemble}) shows that the local physics described
by the transfer function $T_l(\beta)$ is interconnected with
the global structure described by the quadratic sum over the
coefficients $\xi_{lm}^{\beta,i}({\cal M};t)$.
For fixed value of $l$, both contributions depend on the wave number $\beta$
so that neither factor is responsible on its own for a pronouced
suppression of correlations.
Attempts to explain the low correlations with only the first few modes in
$\beta$ turn out to fail since there is no sharp boundary between the modes
determining the large scale behaviour and those that do not.
So that one cannot simply analyse a truncated series instead of
eq.\,(\ref{Eq:Cl_ensemble}).

\begin{figure}
\begin{center}
\vspace*{-80pt}
\begin{minipage}{16cm}
\begin{minipage}{8cm}
\hspace*{20pt}\includegraphics[width=10.0cm]{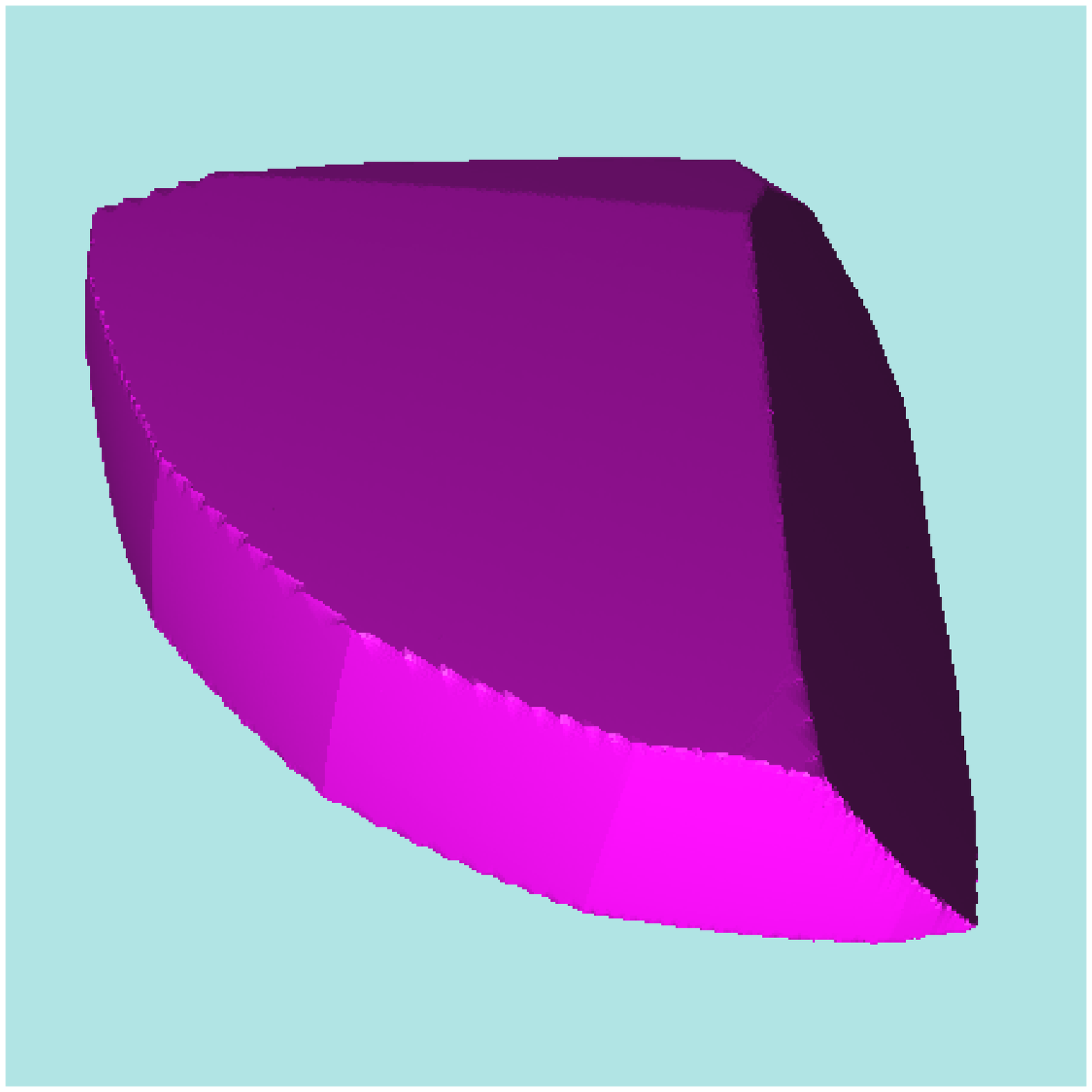}
\put(-220,285){(a)}
\end{minipage}
\begin{minipage}{8cm}
\hspace*{0pt}\includegraphics[width=10.0cm]{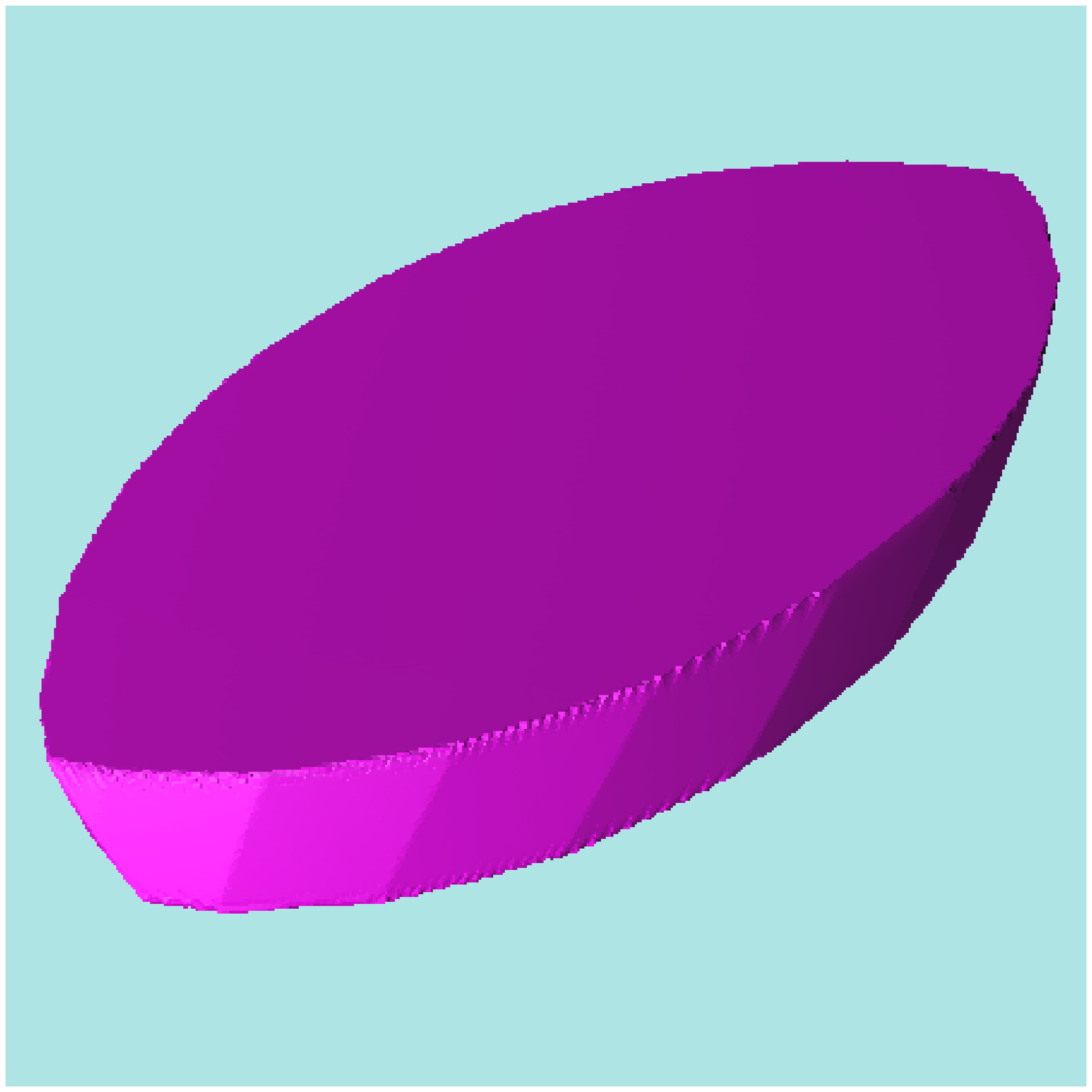}
\put(-220,285){(b)}
\end{minipage}
\end{minipage}
\vspace*{-100pt}
\end{center}
\caption{\label{Fig:Dirichlet_DZ_16_9}
The Dirichlet domains for $DZ(16,9)$ are shown with
$\rho=0.086$ and $\alpha=0$ in panel (a) and with
$\rho=\frac\pi 4$ and $\alpha=0$ in panel (b).
}
\end{figure}

In section \ref{sec:S_statistics} it was shown
that a much stronger observer-position dependence
with respect to the $S$ statistics occurs in $DZ(16,9)$
compared to $DZ(16,3)$.
The $DZ(16,9)$ space possesses a first minimum in the $S$ statistics
at $\Omega_{\hbox{\scriptsize tot}}=1.006$ for the observer at
$\rho=0.086$ and $\alpha=0$.
At this value of $\Omega_{\hbox{\scriptsize tot}}$ the maximal
CMB anisotropy even exceeds that of the simply connected ${\cal S}^3$ space
for an observer at $\rho=\frac\pi 4$ and $\alpha=0$.
Figure \ref{Fig:Dirichlet_DZ_16_9} shows for both positions
the corresponding Dirichlet domains emphasising
that the large variability in the $S$ statistics is also reflected
in the Dirichlet domains.
The $DZ(16,9)$ space cannot have a Dirichlet domain corresponding
to regular polyhedral spaces as shown in figure \ref{Fig:Dirichlet_Polyeder}
since the group orders do not match.


Although deviations from the predictions of the well-proportioned conjecture
are found, it is nevertheless remarkable
that the Dirichlet domains of two regular polyhedral spaces
play a role with respect to the CMB anisotropy suppressions
in double-action manifolds
$DZ(p,n) = {\cal S}^3/\left(D^\star_p \times Z_n\right)$.

\section{Comparison of Prism Double-Action Manifolds with Observations}
\label{sec:I_statistics}

Up to now the power of the correlation function for the manifolds $DZ(p,n)$
on scales larger than $60^{\circ}$ is studied without
reference to observations.
In this section the correlations of the CMB for these models are
compared with that of the WMAP 7yr data on all angular scales 
$\vartheta\in[0^\circ,180^\circ]$ using the correlation function 
of the ILC 7yr map \cite{Gold_et_al_2010}. 
The integrated weighted temperature correlation difference
\cite{Aurich_Janzer_Lustig_Steiner_2007}
\begin{equation}
\label{Eq:I_measure}
I := \int_{-1}^1 \hbox{d}\cos\vartheta \; \;
\frac{(C^{\hbox{\scriptsize model}}(\vartheta)-
C^{\hbox{\scriptsize obs}}(\vartheta))^2}
{\hbox{Var}(C^{\hbox{\scriptsize model}}(\vartheta))}
\hspace{10pt} 
\end{equation}
is suited for such a comparison of the observed correlation function 
$C^{\hbox{\scriptsize obs}}(\vartheta)$ with that of the model 
$C^{\hbox{\scriptsize model}}(\vartheta)$,
where the ensemble average due to the Gaussian initial
condition is used for the latter.
Furthermore, the model is normalised to the angular power spectrum of the
WMAP data using the multipoles between $l=20$ and 45.
The cosmic variance of the model is computed using
\begin{equation}
\label{Eq:Var_C_theta}
\hbox{Var}(C(\vartheta)) \; \approx \;
\sum_l \frac{2l+1}{8\pi^2} \,  \left[C_l \,P_l(\cos\vartheta)\right]^2
\hspace{10pt} .
\end{equation}


\begin{figure}
\begin{center}
\begin{minipage}{10cm}
\vspace*{-25pt}
\hspace*{18pt}\includegraphics[width=10.0cm]{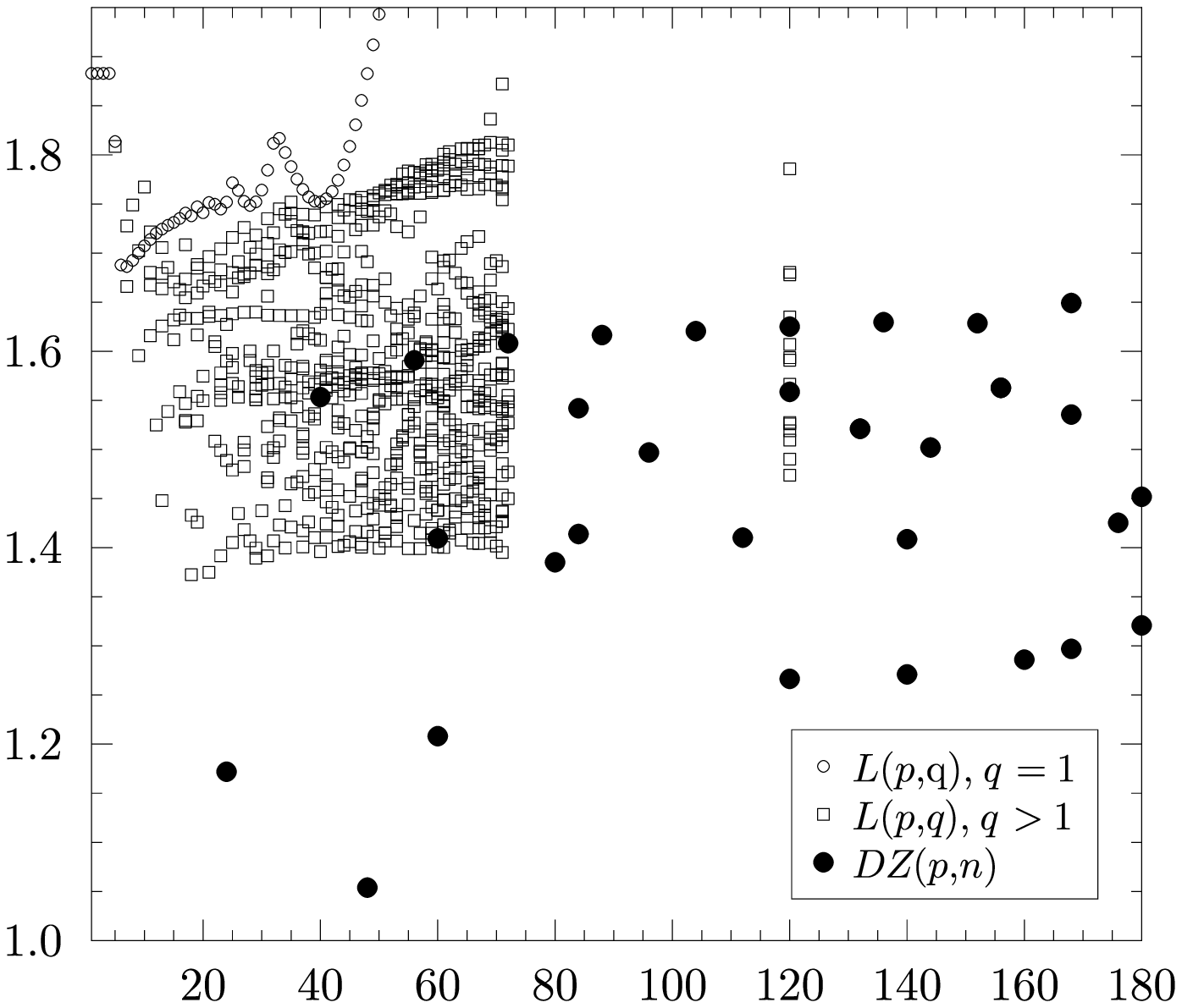}
\put(-130,200){(a) no mask}
\put(-25,28){$|\Gamma|$}
\put(-320,195){$I_{\hbox{\scriptsize min}(\alpha,\rho,\Omega_{\hbox{\scriptsize tot}})}$}
\put(-212,68){\tiny $DZ(8,3)$}
\put(-186,46){\tiny $DZ(16,3)$}
\put(-172,75){\tiny $DZ(20,3)$}
\end{minipage}
\begin{minipage}{10cm}
\vspace*{-50pt}
\hspace*{18pt}\includegraphics[width=10.0cm]{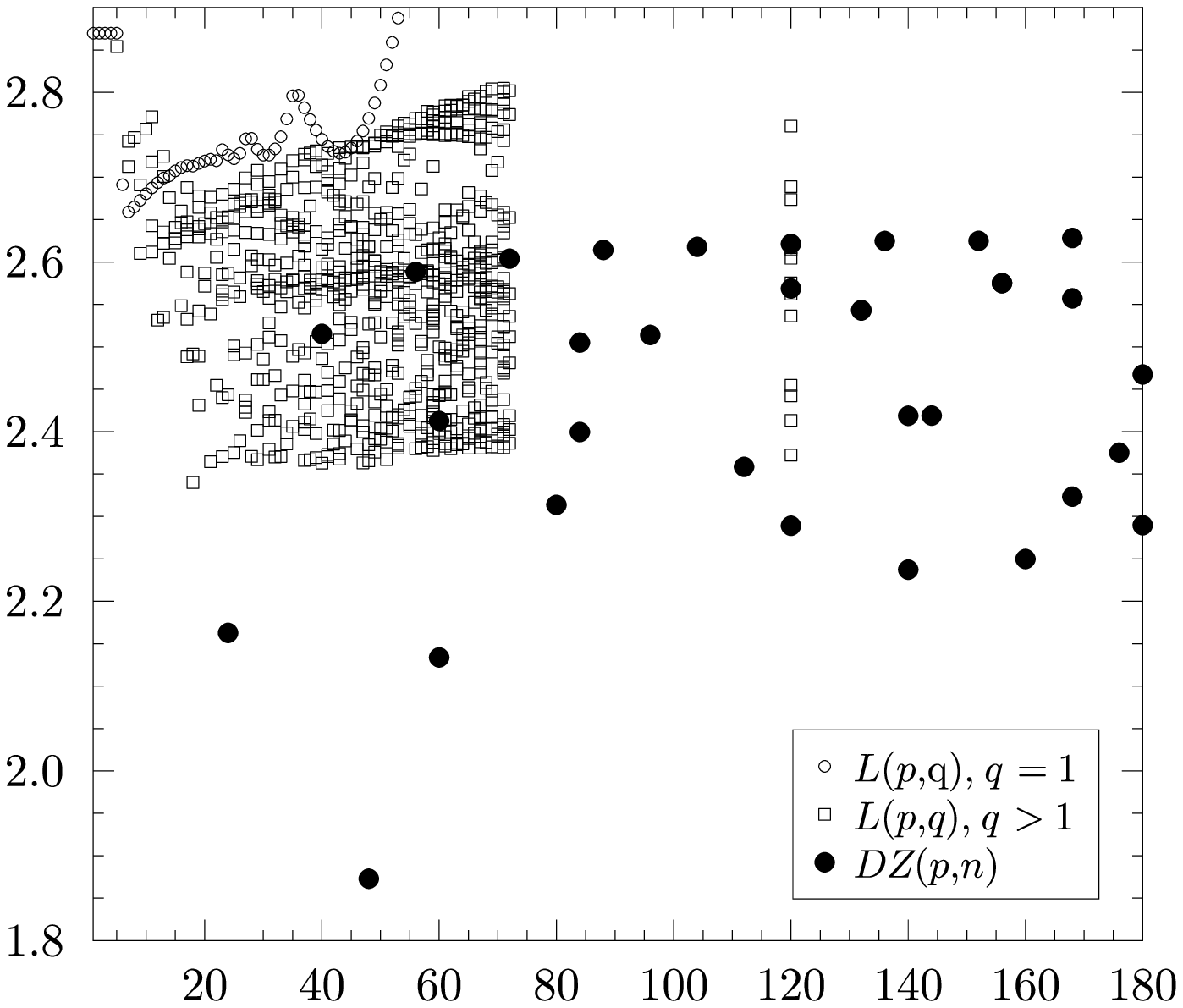}
\put(-130,200){(b) KQ85 mask}
\put(-25,28){$|\Gamma|$}
\put(-320,195){$I_{\hbox{\scriptsize min}(\alpha,\rho,\Omega_{\hbox{\scriptsize tot}})}$}
\put(-212,94){\tiny $DZ(8,3)$}
\put(-186,47.5){\tiny $DZ(16,3)$}
\put(-172,90){\tiny $DZ(20,3)$}
\end{minipage}
\begin{minipage}{10cm}
\vspace*{-50pt}
\hspace*{18pt}\includegraphics[width=10.0cm]{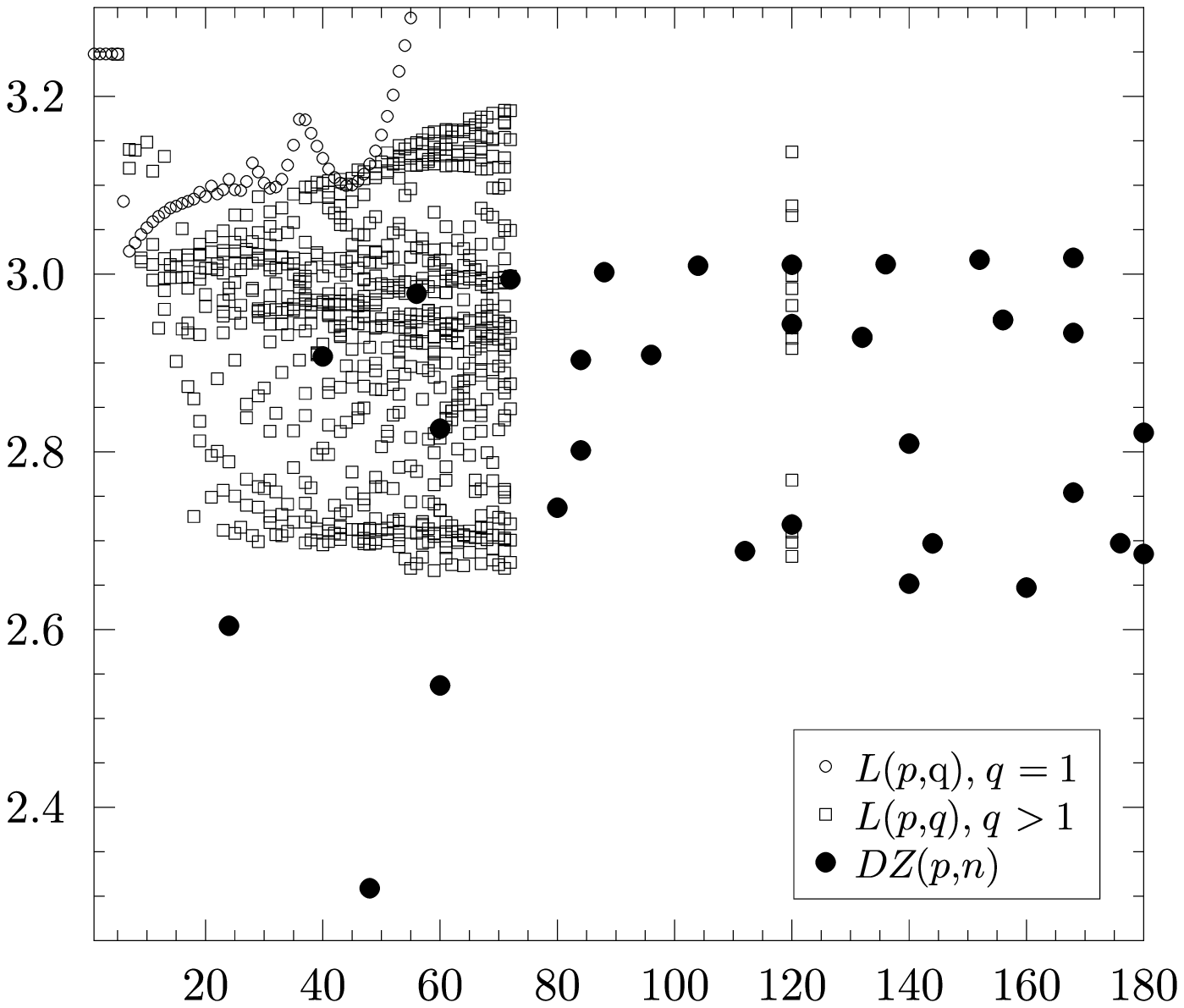}
\put(-130,200){(c) KQ75 mask}
\put(-25,28){$|\Gamma|$}
\put(-320,195){$I_{\hbox{\scriptsize min}(\alpha,\rho,\Omega_{\hbox{\scriptsize tot}})}$}
\put(-212,95.5){\tiny $DZ(8,3)$}
\put(-186,46){\tiny $DZ(16,3)$}
\put(-172,84.5){\tiny $DZ(20,3)$}
\end{minipage}
\vspace*{-25pt}
\end{center}
\caption{\label{Fig:I_statistic}
The minima of the $I$ statistics taken over $\alpha$, $\rho$, and
$\Omega_{\hbox{\scriptsize tot}}$ are plotted for the
double-action manifolds $DZ(p,n)$ as full disks.
The open circles and open boxes represent the
homogeneous and inhomogeneous lens spaces $L(p,q)$.
These are complete for $p\leq 72$ and, in addition,
the group order $p=120$ is also shown.
The $\Omega_{\hbox{\scriptsize tot}}$ interval is restricted
for both space classes to $\Omega_{\hbox{\scriptsize tot}}\in [1.001,1.05]$.
}
\end{figure}


The correlation difference $I$ is calculated for three different 
correlation functions $C^{\hbox{\scriptsize obs}}(\vartheta)$
which are computed in all three cases from the ILC 7yr map of the WMAP data
but are based on different subsets of pixels.
One correlation function $C^{\hbox{\scriptsize obs}}(\vartheta)$
is obtained from the complete ILC 7yr map
but the other two apply the KQ75 7yr and KQ85 7yr masks.
The two masks are provided by \cite{Gold_et_al_2010}, where
the KQ85 7yr and the KQ75 7yr masks include 78.3\% and 70.3\% of the sky, 
respectively.
For a discussion to this topic see e.\,g.\ \cite{Aurich_Lustig_2012a}. 

The values of the $I$ statistics are computed for all 33 prism double-action
manifolds $DZ(p,n)$ up to the group order $|\Gamma|=180$ for the same
values of $\Omega_{\hbox{\scriptsize tot}}$, $\alpha$, and $\rho$
as in the previous sections.
For a given manifold the best value for the $I$ statistics, i.\,e.\
\begin{equation}
\label{Eq:I_min}
I_{\hbox{\scriptsize min}(\alpha,\rho,\Omega_{\hbox{\scriptsize tot}})}
\; = \;
\hbox{min}_{\{ \alpha,\rho,\Omega_{\hbox{\scriptsize tot}}\}} \;
I(\alpha,\rho,\Omega_{\hbox{\scriptsize tot}})
\hspace{10pt} ,
\end{equation}
is then determined where the $\Omega_{\hbox{\scriptsize tot}}$ interval
is restricted to $\Omega_{\hbox{\scriptsize tot}}\in [1.001,1.05]$.
These values are plotted as full disks in figure \ref{Fig:I_statistic}
where the three panels refer to the three different observational
correlation functions $C^{\hbox{\scriptsize obs}}(\vartheta)$.
A survey of lens spaces $L(p,q)$ is provided in \cite{Aurich_Lustig_2012b}
where the $S$ statistics as well as the $I$ statistics are analysed
for all lens spaces $L(p,q)$ up to the group order $p=72$.
It is found that two sequences of lens spaces exist
with a relatively strong CMB anisotropy suppression.
These data are also shown in figure \ref{Fig:I_statistic}
where the correlation difference $I$ of these lens spaces is plotted
as open circles for the homogeneous spaces $L(p,1)$
and as open squares for the inhomogeneous spaces $L(p,q)$, $q>1$.
Three $DZ(p,n)$ spaces immediately attract attention
because their values of the $I$ statistics are significantly smaller
than in any of the lens spaces $L(p,q)$.
These three spaces are $DZ(16,3)$, $DZ(8,3)$, and $DZ(20,3)$.
The double-action manifold $DZ(16,3)$ leads to the best match
for all three observational correlation functions
$C^{\hbox{\scriptsize obs}}(\vartheta)$,
that is with the KQ75 7yr or KQ85 7yr mask or without a mask at all.
The two other spaces change the second and third position
with respect to the lowest value of
$I_{\hbox{\scriptsize min}(\alpha,\rho,\Omega_{\hbox{\scriptsize tot}})}$
depending on the selected pixels of the ILC map.
Using the KQ75 7yr or KQ85 7yr mask the space $DZ(20,3)$
provides a slightly better description of the CMB data than $DZ(8,3)$.
This relation reverses if no mask is applied.

This comparison reveals that the three double-action manifolds
$DZ(16,3)$, $DZ(8,3)$, and $DZ(20,3)$ produce the lowest
CMB correlations among the two classes of inhomogeneous spaces 
$DZ(p,n)$ and $L(p,q)$.


\section{Summary and Discussion}


This paper studies whether a special class of spherical topologies
can alleviate the apparent disagreement between
the standard concordance model of cosmology and measurements of the
low power at large scales in the cosmic microwave background maps.
The cosmological observations point to an almost spatially flat cosmos.
Thus, we restrict the analysis to spherical models
that are almost flat by confining the total density parameter
$\Omega_{\hbox{\scriptsize tot}}$ to the interval
$\Omega_{\hbox{\scriptsize tot}}=1.001,\dots,1.05$.
The space of such cosmological models is the simply connected 3-sphere
${\cal S}^3$ if the model should be close to the concordance model.
These concordance-like models produce, however, more correlations
in the CMB temperature fluctuations on large angular scales
as observed in CMB sky.
This disagreement can be alleviated by considering instead of the
simply connected 3-sphere ${\cal S}^3$ a multiconnected space
${\cal M}={\cal S}^3/\Gamma$ which is obtained from the 3-sphere ${\cal S}^3$
by tessellating it under the action of a deck group $\Gamma$.

Multiconnected spherical spaces can be classified into several groups,
see e.\,g.\ \cite{Gausmann_Lehoucq_Luminet_Uzan_Weeks_2001}.
On the one hand there are three deck groups
which lead to regular polyhedral spaces.
These are obtained by applying the binary tetrahedral group $T^\star$,
the binary octahedral group $O^\star$, or the binary icosahedral group $I^\star$
to the 3-sphere ${\cal S}^3$.
The polyhedral spaces are homogeneous in the sense that their
fundamental domains defined as a Dirichlet domain look the same
independent of the position of the CMB observer.
The important implication thereof is that the statistical CMB properties
do not depend on the CMB observer.
Inhomogeneous manifolds possess such a position dependence, in general.
It turns out that the regular polyhedral spaces yield CMB anisotropies
with a normalised $S$ statistics, eq.\,(\ref{Eq:S_statistic_60}),
of $S/S_{{\cal S}^3}\simeq 0.11$.
This is the strongest suppression of CMB correlations on large angles
found so far.

In addition to these three manifolds there are the so-called prism spaces,
which are generated by the binary dihedral group $D^\star_p$.
The prism spaces are homogeneous and are analysed up to the
group order $p=72$ in \cite{Aurich_Lustig_2012a},
see also \cite{Lustig_2007}.
The CMB suppression of some prism spaces can be as low as
$S/S_{{\cal S}^3}\simeq 0.8\dots 0.9$ for
$\Omega_{\hbox{\scriptsize tot}}=1.001,\dots,1.05$
as shown in table \ref{Tab:dieder}.
The prism space $D_{12}$ drops even to $S/S_{{\cal S}^3}\simeq 0.68$
at $\Omega_{\hbox{\scriptsize tot}}=1.05$
being the boundary of the imposed $\Omega_{\hbox{\scriptsize tot}}$ interval.
This shows that the regular polyhedral spaces suppress the CMB anisotropies
on large angles stronger than the prism spaces.

A further class of spherical spaces are the lens spaces $L(p,q)$
which are homogeneous for $q=1$ and inhomogeneous for $q>1$.
The CMB properties of the lens spaces are analysed systematically
in the survey presented in \cite{Aurich_Lustig_2012b},
and some interesting sequences of lens spaces $L(p,q)$ are found.
But their CMB suppression is less pronounced
compared to the regular polyhedral spaces.
Typical values are in the range $S/S_{{\cal S}^3}\simeq 0.5\dots 0.6$
\cite{Aurich_Lustig_2012b}.

Therefore, the question emerges whether other classes of multiconnected
spherical spaces can do it better.
The next class is given by the so-called double-action manifolds
where the group elements are composed of a right- and a left-handed
Clifford transformation belonging to homogeneous deck groups $R$ and $L$. 
Choosing $L=Z_n$ as the cyclic subgroup of Clifford translations
and the subgroup $R$ as the binary dihedral group $D^\star_p$,
leads to the prism double-action manifolds 
$DZ(p,n)={\cal S}^3/\left(D^\star_p \times Z_n\right)$
to which this paper is devoted.
Three further classes of double-action manifolds can be generated by
the binary polyhedral groups leading to
$TZ(24,n):={\cal S}^3/\left(T^\star \times Z_n\right)$,
$OZ(48,n):={\cal S}^3/\left(O^\star \times Z_n\right)$, and
$IZ(120,n):={\cal S}^3/\left(I^\star \times Z_n\right)$.
These three groups do not possess analytical expressions
for their eigenmodes in terms of Wigner polynomials
and require a separate numerical treatment.
Their analysis is reserved to another paper.

The CMB suppression of the double-action manifolds $DZ(p,n)$ is compared
with that of the lens spaces $L(p,q)$ in figure \ref{Fig:I_statistic}.
Three spaces attract attention because they reveal a stronger
CMB suppression than all studied lens and prism spaces.
These are the prism double-action manifolds $DZ(16,3)$, $DZ(8,3)$,
and $DZ(20,3)$. 
The smallest large-angle correlations are seen in $DZ(16,3)$
at $\Omega_{\hbox{\scriptsize tot}}=1.036$,
where $S/S_{{\cal S}^3}\simeq 0.291$ is reached.
If one insists on the density interval 
$0.99 < \Omega_{\hbox{\scriptsize tot}} < 1.02$ (95\% CL),
the space $DZ(20,3)$ provides the most interesting
prism double-action manifold
since it has at $\Omega_{\hbox{\scriptsize tot}}=1.02$ a suppression of
$S/S_{{\cal S}^3}\simeq 0.419$.
Although this suppression is remarkable,
it is nevertheless less pronounced than that found in
the regular polyhedral spaces.

Besides the three classes $TZ(24,n)$, $OZ(48,n)$, and $IZ(120,n)$,
which are not studied in this paper,
there are linked action manifolds
\cite{Gausmann_Lehoucq_Luminet_Uzan_Weeks_2001}
that are not analysed with respect to their CMB suppression until now.
Except for these cases,
one can summarise that the regular polyhedral spaces
are the most promising spherical spaces with respect to their CMB suppression
followed by some members of the class of double-action manifolds $DZ(p,n)$,
notably the spaces $DZ(16,3)$, $DZ(8,3)$, and $DZ(20,3)$.



\section*{Acknowledgements}

We would like to thank the Deutsche Forschungsgemeinschaft
for financial support (AU 169/1-1).
The WMAP data from the LAMBDA website (lambda.gsfc.nasa.gov)
were used in this work.


\section*{References}

\bibliography{../bib_astro}

\bibliographystyle{h-physrev5}

\end{document}